\documentclass[twocolumn,aps,pre]{revtex4-1}
\usepackage{amsmath,amsfonts,amssymb,amsthm}
\usepackage[T1]{fontenc}
\usepackage{bbold}
\usepackage{bm}        % for math
\usepackage{graphicx}  % needed for figures
\usepackage{color} %ADDED FOR QUERIES
\usepackage{stackrel}%ADDED TO IMPROVE APPEARANCE OF EQS. (21) AND (29)

\newtheoremstyle{query}%
{}{}%space above/below
{\color{red}}%body style
{}%heading indent
{\sffamily\bfseries}{:}{12pt}%heading style/punctuation/space after
{}% head spec
\theoremstyle{query}

\begin{document}

\title{The Stretch to Stray on Time: Resonant Length of Random Walks in a Transient}

\author{Martin Falcke$^{1,2,}$}
\email[Martin Falcke: ]{Martin.Falcke@mdc-berlin.de}
%\affiliation{Max Delbr\"uck Center for Molecular Medicine, Robert R\"ossle Str. 10, 13125 Berlin, Germany}
\author{V. Nicolai Friedhoff$^{1}$}
%\email[V. Nicolai Friedhoff: ]{Nicolai.Friedhoff@mdc-berlin.de}
\affiliation{$^1$Max Delbr\"uck Center for Molecular Medicine, Robert R\"ossle Str. 10, 13125 Berlin, Germany\\$^2$Dept. of Physics, Humboldt University, Newtonstr. 15, 12489 Berlin, Germany}
\date{\today}
\begin{abstract}
First-passage times in random walks have a vast number of diverse applications in physics, chemistry, biology, and finance. In general, environmental conditions for a stochastic process are not constant on the time scale of the average first-passage time, or control might be applied to reduce noise. We investigate moments of the first-passage time distribution under a transient describing relaxation of environmental conditions. We solve the Laplace-transformed (generalized) master equation analytically using a novel method that is applicable to general state schemes. The first-passage time from one end to the other of a linear chain of states is our application for the solutions.  The dependence of its average on the relaxation rate obeys a power law for slow transients. The exponent $\nu$ depends on the chain length $N$ like $\nu=-N/(N+1)$ to leading order. Slow transients substantially reduce the noise of first-passage times expressed as the coefficient of variation (CV), even if the average first-passage time is much longer than the transient. The CV has a pronounced minimum for some lengths, which we call resonant lengths. These results also suggest a simple and efficient noise control strategy, and are closely related to the timing of repetitive excitations, coherence resonance and information transmission by noisy excitable systems. A resonant number of steps from the inhibited state to the excitation threshold and slow recovery from negative feedback provide optimal timing noise reduction and information transmission.
\end{abstract}

%\pacs{05.40.-a, 05.40.Fb, 05.40.Jc, 05.20.–y, 05.10.Gg, 05.60.-k}
\maketitle

%\tableofcontents

\section{Introduction}

Continuous-time random walks are a unifying concept across  physics~\cite{bechinger2000,Voituriez2015,voituriez2016,Molini2011,panja2010,Turiv2013,jacquin2014,weiss2012,baranowskii2014,metzler2000,blumen2002,Ivanyuk2003,sokolov2005,shlesinger2006}, chemistry~\cite{VanKampen:01,oppenheim77,Voituriez2010}, biology~\cite{Berg93,gazzola2009,bressloff2013,pal2016}, and finance~\cite{scott97,steele2000}. The drunkard's straying on his way home from the pub is the graphic example frequently used to illustrate the randomness of step timing and direction. In particular, the time of first passage of a specific state given the stochastic nature of the process is of interest in many applications. It describes the time the drunkard arrives home, the time necessary for a chemical reaction or gene expression to reach a certain number of product molecules~\cite{VanKampen:01,pal2016}, or a quantity in one of the many other applications~\cite{Voituriez2015,voituriez2016,Ivanyuk2003,Voituriez2010,Mayo2011,Winter1981,Berg1981,Riggs1970,Riggs1970a,Shimamoto1999,buttiker2012,Lax73,britton2010,RednerMetzlerOshanin,Redner,klafter2007,shlesinger2007,elowitz2017a}.

While noise happens on the small length scales and short time scales of a system, it may trigger events on a global scale. One of the most important functions of noise for macroscopic dynamics arises from its combination with thresholds~\cite{Schimansky2000a,schimansky2003a,Schimansky2004d}. These are defined by the observation that the dynamics of a system remains close to a stable state as long as it does not cross the threshold value, and an actively amplified deviation from this state happens when the threshold is crossed. Noise drives the system  across the threshold in a random manner. First passage is a natural concept to describe the timing of threshold crossings. Ignition processes are an illustrative textbook example. Although a small random spark might not be capable of igniting an inflammable material,  a few of them might cause an explosion or forest fire. If the system again attains  its stable state upon recovery from a deviation, such behavior is called excitable and the large deviation an excitation. The excitation is terminated by negative feedback. A forest is excitable, because it regrows after a fire. Consumption of inflammable trees acts as the negative feedback. Excitability describes not only forest fires but also the dynamics of heterogeneous catalysis~\cite{falcke92a}, the firing of neurons~\cite{Keener98}, the properties of heart muscle tissue~\cite{Keener98}, and many other systems in physics, chemistry, and biology~\cite{scott75,Moss95,CrossHohenberg93,Mikhailov,Ebeling86}.

Random walks are frequently defined on a set of discrete states. The rates $f_{i,j}$ or waiting-time distributions $\Psi_{i,j}$ for transitions from state $i$ to $j$ set the state dwell times. The first-passage time between two widely separated states is much longer than the individual dwell times, and the conditions setting the rates and parameters of the $\Psi_{i,j}$ are likely to change or external control acts on the system between start and first passage. The conditions for igniting a forest fire change with the seasons or  because the forest recovers from a previous fire. The occurrence of subcritical sparks during  recovery has essentially no affect on the process of regrowth. More generally,  noise does not affect  recovery on large space and long time scales, and the random process experiences  recovery as a slow deterministic change of environmental conditions. Since recovery is typically a slow relaxation process~\cite{CrossHohenberg93,Mikhailov,Ebeling86}, it dominates event timing. Hence, first passage in an exponential transient is a natural concept through which to understand the timing of sequences of excitations. We will investigate it in this study.

We will take Markovian processes as one of the asymptotic cases of  transient relaxation. There are several reasons for also considering  non-Markovian formulations of continuous-time random walks. A description of a diffusing particle becomes non-Markovian whenever the  particle  not only passively experiences the thermal fluctuations causing its diffusive motion, but also acts back on its immediate surroundings~\cite{bechinger2000,voituriez2016,franosch2011,panja2010,Turiv2013,jacquin2014,weiss2012}. Non-Markovian waiting-time distributions for state transitions arise naturally in transport theory~\cite{Lindenberg2008,buttiker2012,Albert2016,Oppen2005}. Frequently in biological applications,  the discrete states that arise are lumped states consisting of many ``microscopic'' states. Groups of open and closed states of ion channels lumped into single states are examples of this~\cite{Hille,falcke2005c_Sneyd_comparison}. Transitions between lumped states are non-Markovian owing to the internal dynamics. We may also use waiting-time distributions if we lack information on all the individual steps of a process, but we do know the inter-event interval distributions. This is usually the case with \emph{in vivo} measurements such as stimulation of a cell and the appearance of a response, or differentiation sequences of stem cells~\cite{elowitz2017a}. The state probabilities of non-Markovian processes obey generalized master equations, which we will use here~\cite{oppenheim77,kenkre1978,talkner85,west86,metzler2000,metzler99}.

In Sec.~\ref{sec:basicequations}, we  present a formulation of the general problem in terms of the normal and generalized master equations and give analytic solutions for both of these. These solutions apply to general state schemes. We continue with investigating first passage on linear chains of states in  Sec.~\ref{sec:FPT}. We present results on scaling of the average first-passage time with the relaxation rate of the transient $\gamma$ and the chain length $N$ in Sec.~\ref{sec:scaling}, and results on the phenomenon of resonant lengths in Sec.~\ref{sec:resonant}.

\section{Basic equations}\label{sec:basicequations}
\subsection{The asymptotically Markovian master equation}
In this section, we consider transition rates relaxing with rate $\gamma$ to an asymptotic value $\lambda_{i,j}$ like
\begin{equation}\label{eq:McaseB}
  f_{i,j}(t)=\lambda_{i,j}\left(1+B_{i,j}e^{-\gamma t}\right),\ \ \lambda_{i,j}\ge 0,\ B_{i,j}\ge -1.
\end{equation}
They reach a Markov process asymptotically. The dynamics of the probability $P_{i,j}(t)$ to be in state $j$ for a process that started in $i$ at $t=0$ obey the master equation
\begin{eqnarray}\label{eq:MasterEquation}
   \frac{dP_{i,j}}{dt} &=& \sum_{k=0}^{N}\lambda_{k,j} P_{i,k} -\lambda_{j,k} P_{i,j}
  \nonumber\\
  &&+ e^{-\gamma t}\left(\lambda_{k,j}B_{k,j}P_{i,k}  -\lambda_{j,k}B_{j,k}P_{i,j}\right).
\end{eqnarray}
In matrix notation with the vector of probabilities $P_i$, we have
\begin{eqnarray}\label{eq:MatrixMasterEquation}
  \frac{dP_{i}}{dt} &=& E P_{i} + e^{-\gamma t} D P_{i},
\end{eqnarray}
with the matrices $E$ and $D$ defined by Eq.~\eqref{eq:MasterEquation}. The initial condition defines the vector $r_i=\{\delta_{ij}\}, j=0,\ldots,N$. The Laplace transform of the master equation allows for a comfortable calculation of moments of the first-passage times, which we will carry out in Sec.~\ref{sec:FPT}. The Laplace transform of Eq.~\eqref{eq:MatrixMasterEquation} is the system of linear difference equations
\begin{eqnarray}\label{eq:LPMasterEquation}
  s\tilde{P}_{i}(s) - r_i &=& E \tilde{P}_{i}(s) + D \tilde{P}_{i}(s+\gamma).
\end{eqnarray}

\subsection{The generalized master equation}
\subsubsection{The waiting-time distributions}
Waiting-time distributions $\Psi_{j,k}$ in a constant environment depend on the time $t-t'$ elapsed since the process entered state $j$ at time $t'$. The change in conditions causes an additional dependence on $t$: $\Psi_{j,k}(t,t-t')$. The lumping of states, which we introduced as a major cause of dwell-time-dependent transition probabilities, often entails $\Psi_{j,k}\left(t,0\right)=\Psi_{j,k}\left(t,\infty\right)=0$, with a maximum of $\Psi_{j,k}\left(t,t-t'\right)$ at intermediate values of $t-t'$. Waiting-time distributions used in transport theory exhibit similar properties~\cite{Wilkins92,Lindenberg2008,buttiker2012,Albert2016}.  We use a simple realization of this type of distributions by a biexponential function in $t-t'$:
\begin{eqnarray}%\label{eq:Psipm1}
\Psi_{j,k}\left(t,t-t'\right)&=&A_{j,k}\big(e^{-\alpha_{j,k}(t-t')}- e^{-\beta_{j,k}(t-t')}\big)\nonumber\\&&\times\left(1+M_{j,k}e^{-\gamma t}\right)\label{eq:Psijk}\label{eq:ghdefinitionfirstrow}\\
&=& g_{j,k}(t-t') + h_{j,k}(t-t') e^{-\gamma t}.\label{eq:ghdefinition}
\end{eqnarray}
The transient parts $h_{j,k}(t-t')e^{-\gamma t}$ of $\Psi_{j,k}\left(t,t-t'\right)$ collect all factors of $e^{-\gamma t}$ in Eq.~\eqref{eq:ghdefinitionfirstrow}. The functions $g_{j,k}(t-t')$ describe the asymptotic part of the waiting-time distributions remaining after the transient. The $\Psi_{j,k}(t,t-t')$ are normalized to the splitting probabilities $C_{j,k}=\int_{t'}^\infty dt\, \Psi_{j,k}\left(t,t-t'\right)$ (the total probability for a transition from $j$ to $k$ given the system entered $j$ at $t'$). They satisfy
\begin{eqnarray}\label{eq:Splitnorm}
\sum_{k=1}^{N_\mathrm{out}^{(j)}}C_{j,i_k}(t')&=&1-C_{j,j}(t'),
\end{eqnarray}
where $N_\mathrm{out}^{(j)}$ is the number of transitions out of state $j$.

\subsubsection{The generalized master equation and its Laplace transform}
In the non-Markovian case, the dynamics of the probabilities $P_{i,j}(t)$ obey a generalized master equation~\cite{kenkre1973,montroll73,Burshtein1986,sokolov2005a,sokolov2007}
\begin{equation}\label{eq:MasterIntDiff}
    \frac{dP_{i,j}(t)}{dt} = \sum_{l=1}^{N_\mathrm{in}^{(j)}} I^i_{k_l,j}(t) - \sum_{l=1}^{N_\mathrm{out}^{(j)}} I^i_{j,i_l}(t),
\end{equation}
where $I^i_{l,j}(t)$ is the probability flux due to transitions from state $l$ to $j$ given that the process started at state $i$ at $t=0$, and $N_\mathrm{in}^{(j)}$ is the number of transitions toward $j$. The fluxes are the solutions of the integral equation
\begin{equation}\label{eq:probfluxes}
I_{l,j}^i(t) = \int_0^t  dt'\, \Psi_{l,j}(t,t-t') \sum_{k=1}^{N_\mathrm{in}^{(l)}}I_{i_k,l}^i(t') + q_{l,j}^i(t).
\end{equation}
The second factor in the convolution is the probability of arriving in state $l$ at time $t'$, and the first factor is the probability of leaving toward $j$ at time $t$ given arrival at $t'$. The $q_{l,j}^i$ are the initial fluxes, with $q_{l,j}^i(t)\equiv0$ for $i\ne l$.

The Laplace transform of the probability dynamics equation \eqref{eq:MasterIntDiff} is
\begin{equation}\label{eq:laplaceME}
s \tilde P_{i,j}(s) - \delta_{ij} = \sum_{l=1}^{N_\mathrm{in}^{(j)}}\tilde{I}_{k_l,j}^i(s) - \sum_{l=1}^{N_\mathrm{out}^{(j)}}\tilde{I}_{j,i_l}^i(s),
\end{equation}
which contains the Laplace-transformed probability fluxes $\tilde{I}_{l,j}^i$. The Kronecker delta $\delta_{ij}$ captures the initial condition.

Laplace-transforming Eq.~\eqref{eq:probfluxes} is straightforward for terms containing the asymptotic part $g_{l,j}(t-t')$ of the $\Psi_{l,j}\left(t,t-t'\right)$ [see Eq.~\eqref{eq:ghdefinition}], since they depend on $t-t'$ only and the convolution theorem applies directly.  The terms containing the transient part $h_{l,j}(t-t') e^{-\gamma t}$ depend on both  $t-t'$ and $t$ and require a little more attention:
\begin{eqnarray}
\lefteqn{\int_{0}^{\infty} dt\, e^{-s t}\int_{0}^{t} dt'\, h_{l,j}(t-t') e^{-\gamma t}\sum_{k=1}^{N_\mathrm{in}^{(l)}}I_{i_k,l}^i(t')}&&\nonumber\\
&=&\int_{0}^{\infty} dt\, e^{-(s+\gamma)t}\int_{0}^{t} dt'\, h_{l,j}(t-t')\sum_{k=1}^{N_\mathrm{in}^{(l)}}I_{i_k,l}^i(t')\nonumber\\
&=&\tilde{h}_{l,j}(s+\gamma)\sum_{k=1}^{N_\mathrm{in}^{(l)}}\tilde{I}_{i_k,l}^i(s+\gamma).
\end{eqnarray}
This leads to the Laplace transform of Eq.~\eqref{eq:probfluxes}:
\begin{eqnarray}\label{eq:probfluxeslaplace}
\tilde{I}_{l,j}^i(s) &=& \tilde{q}_{l,j}^i(s) + \tilde{g}_{l,j}(s)\sum_{k=1}^{N_\mathrm{in}^{(l)}}\tilde{I}_{i_k,l}^i(s)\nonumber \\
&&+\tilde{h}_{l,j}(s+\gamma)\sum_{k=1}^{N_\mathrm{in}^{(l)}}\tilde{I}_{i_k,l}^i(s+\gamma).
\end{eqnarray}
We write the $\tilde{I}_{l,j}^i(s)$ and $\tilde{q}_{l,j}^i(s)$ as the vectors $\tilde{I}^i(s)$ and $\tilde{q}^i(s)$, respectively, to obtain
\begin{eqnarray}
\tilde{I}^i(s) &=& \tilde{G}(s)\tilde{I}^i(s)+\tilde{H}(s+\gamma)\tilde{I}^i(s+\gamma) + \tilde{q}^i(s).\label{eq:probfluxeslaplace0}
\end{eqnarray}
Solving for $\tilde{I}^i(s)$ results in
\begin{equation}
\tilde{I}^i(s)=\big[\mathbb{1}-\tilde{G}(s)\big]^{-1}\big[\tilde{H}(s+\gamma)\tilde{I}^i(s+\gamma) + \tilde{q}^i(s)\big].\label{eq:probfluxeslaplace1}
\end{equation}
This is again a system of linear difference equations. The entries in the matrices $\tilde{G}(s)$ and $\tilde{H}(s)$ are the functions $\tilde{g}_{l,j}(s)$ and $\tilde{h}_{l,j}(s)$, which are the Laplace transforms of $g_{l,j}(t-t')$ and $h_{l,j}(t-t')$ [Eq.~\eqref{eq:ghdefinition}].

\subsubsection{Solving the Laplace-transformed generalized and asymptotically Markovian master equations}

All elements of $\tilde{H}(s)$ [Eq.~\eqref{eq:probfluxeslaplace1}] vanish for $s\rightarrow\infty$. We also expect $\tilde{I}^i(s)$ to be bounded for $s\rightarrow\infty$. Hence, the solution for $\gamma=\infty$ is
\begin{eqnarray}\label{eq:inhomogeneous}
\tilde{I}_{\infty}^i(s) &=& \big[\mathbb{1}-\tilde{G}(s)\big]^{-1}\tilde{q}^i(s).
\end{eqnarray}
Consequently, for $k\gamma\gg$1,
\begin{eqnarray}\label{eq:keins}
\tilde{I}^i(s+k\gamma) &\approx& \big[\mathbb{1}-\tilde{G}(s+k\gamma)\big]^{-1}\tilde{q}^i(s+k\gamma)\nonumber
\end{eqnarray}
holds for the solution $\tilde{I}^i(s)$ for all values of $\gamma>0$ and for $k$ a natural number. Once we know $\tilde{I}^i(s+k\gamma)$, we can use Eq.~\eqref{eq:probfluxeslaplace1} to find $\tilde{I}^i(s+(k-1)\gamma)$:
\begin{multline}\label{eq:kminuseins}
\tilde{I}^i(s+(k-1)\gamma) \approx \big[\mathbb{1}-\tilde{G}(s+(k-1)\gamma)\big]^{-1}\\
\times\big\{\tilde{H}(s+k\gamma)\big[\mathbb{1}-\tilde{G}(s+k\gamma)\big]^{-1}\tilde{q}^i(s+k\gamma)\\
+\tilde{q}^i(s+(k-1)\gamma)\big\}, \ \ \ k\gg\gamma^{-1}.
\end{multline}
In this way, we can consecutively use Eqs.~\eqref{eq:probfluxeslaplace1} and \eqref{eq:inhomogeneous} to express $\tilde{I}^i(s+(k-j)\gamma)$, $j=0,\ldots,k$, by known functions. The solution becomes exact with $k\rightarrow\infty$. With the definition $\tilde{A}(s)=\big[\mathbb{1}-\tilde{G}(s)\big]^{-1}\tilde{H}(s+\gamma)$, we obtain
\begin{multline}\label{eq:Iinh}
\tilde{I}^i(s)=\big[\mathbb{1}-\tilde{G}(s)\big]^{-1}\tilde{q}^i(s)\\
 +\sum_{k=1}^\infty \prod_{j=0}^{k-1} \tilde{A}(s+j\gamma)
  \big[\mathbb{1}-\tilde{G}(s+k\gamma)\big]^{-1}\tilde{q}^i(s+k\gamma)
\end{multline}
as the solution of  Eq.~\eqref{eq:probfluxeslaplace1}. Equation~\eqref{eq:Iinh} is confirmed by verifying that it provides the correct solution in the limiting cases $\gamma=0$ and $\gamma=\infty$. In the latter, all entries of the matrix $\tilde{H}(s)$ vanish, and we obtain directly the result without transient, Eq.~\eqref{eq:inhomogeneous}. For $\gamma=0$, we notice that
\begin{eqnarray*}
\sum_{k=1}^\infty \prod_{j=0}^{k-1} \tilde{A}(s)&=&\big\{\mathbb{1}-\big[\mathbb{1}-\tilde{G}(s)\big]^{-1}\tilde{H}(s)\big\}^{-1}-\mathbb{1}
\end{eqnarray*}
holds, which leads to the correct solution of  Eq.~\eqref{eq:probfluxeslaplace1}:
\begin{eqnarray}
  \tilde{I}^i(s)
  &=&\big[\mathbb{1}-\tilde{G}(s)-\tilde{H}(s)\big]^{-1}\tilde{q}^i(s).
\end{eqnarray}

We now turn to the asymptotically Markovian master equation~\eqref{eq:MasterEquation} and write its Laplace transform \eqref{eq:LPMasterEquation} as
\begin{eqnarray}\label{eq:MasterMarkLaplace}
  \tilde{P}_i(s) &=& \left(\mathbb{1}s-E\right)^{-1}\big[D\tilde{P}_i(s+\gamma) + r_i\big].
\end{eqnarray}
This equation has the same structure as Eq.~\eqref{eq:probfluxeslaplace1}, and since the matrix $\left(\mathbb{1}s-E\right)^{-1}$ also vanishes for $s\rightarrow\infty$, we can calculate the Laplace transform of the $P_{i,j}$ completely analogously to the non-Markovian case. We define $\tilde{B}(s)=\left(\mathbb{1}s-E\right)^{-1}D$ and obtain
\begin{multline}\label{eq:Marksolution}
\tilde{P}_i(s)=\left(\mathbb{1}s-E\right)^{-1}r_i\\
  +\sum_{k=1}^\infty \prod_{j=0}^{k-1} \tilde{B}(s+j\gamma)
 \big[\mathbb{1}(s+k\gamma)-E\big]^{-1}r_i
\end{multline}
as the solution of Eq.~\eqref{eq:MasterMarkLaplace}.

These solutions for the Laplace transforms of the generalized and asymptotically Markovian master equations with a transient, Eqs.~\eqref{eq:probfluxeslaplace1} and~\eqref{eq:Marksolution}, are not restricted to state-independent waiting-time distributions or rates. They also apply to random walks with space- or state-dependent waiting-time distributions and to arbitrary state networks.

\section{The probability density of the first-passage time  for a linear chain of states}\label{sec:FPT}

A large class of stochastic processes, including all the examples mentioned in the introduction, are represented by  state schemes like
\begin{equation}\label{scheme}
0\stackrel[\Psi_{1,0}]{\Psi_{0,1}}{\rightleftarrows}1\stackrel[\Psi_{2,1}]{\Psi_{1,2}}{\rightleftarrows}2 \ \ldots \ N-1\stackrel[\Psi_{N,N-1}]{\Psi_{N-1,N}}{\rightleftarrows}N,
\end{equation}
%\begin{equation}
%0\
%\begin{matrix}
  % \Psi_{0,1} \\
  %\rightleftarrows \\
  %\Psi_{1,0} \\
%\end{matrix}
%\ 1\
%\begin{matrix}
 % \Psi_{1,2} \\
 % \rightleftarrows \\
 % \Psi_{2,1} \\
%\end{matrix}
%\  2\ \ \ldots \  N-1\
%\begin{matrix}
%  \Psi_{N-1,N} \\
  %\rightleftarrows \\
  %\Psi_{N,N-1} \\
%\end{matrix}
%\ N,
%\end{equation}
which we consider from now on. The definition of the transition probabilities for both the asymptotically Markovian system and the non-Markovian system completes its specification. We use the asymptotically Markovian rates
\begin{eqnarray}\label{eq:Mcase1}
  f_{i,i+1}(t)=\lambda\left(1-e^{-\gamma t}\right),\ \ f_{i,i-1}(t)&=&\lambda
\end{eqnarray}
corresponding to $\lambda_{i,j}=\lambda$, $B_{i,i+1}=-1$, and $B_{i,i-1}=0$ in Eq.~\eqref{eq:McaseB}. The process has a strong initial bias for motion toward 0, $C_{i,i-1}(t'=0)>C_{i,i+1}(t'=0)$. It relaxes with rate $\gamma$ to a symmetric random walk with $C_{i,i-1}(t'=\infty)=C_{i,i+1}(t'=\infty)=\frac 12$.

\begin{figure}
  \centering
\includegraphics[width=.5\linewidth]{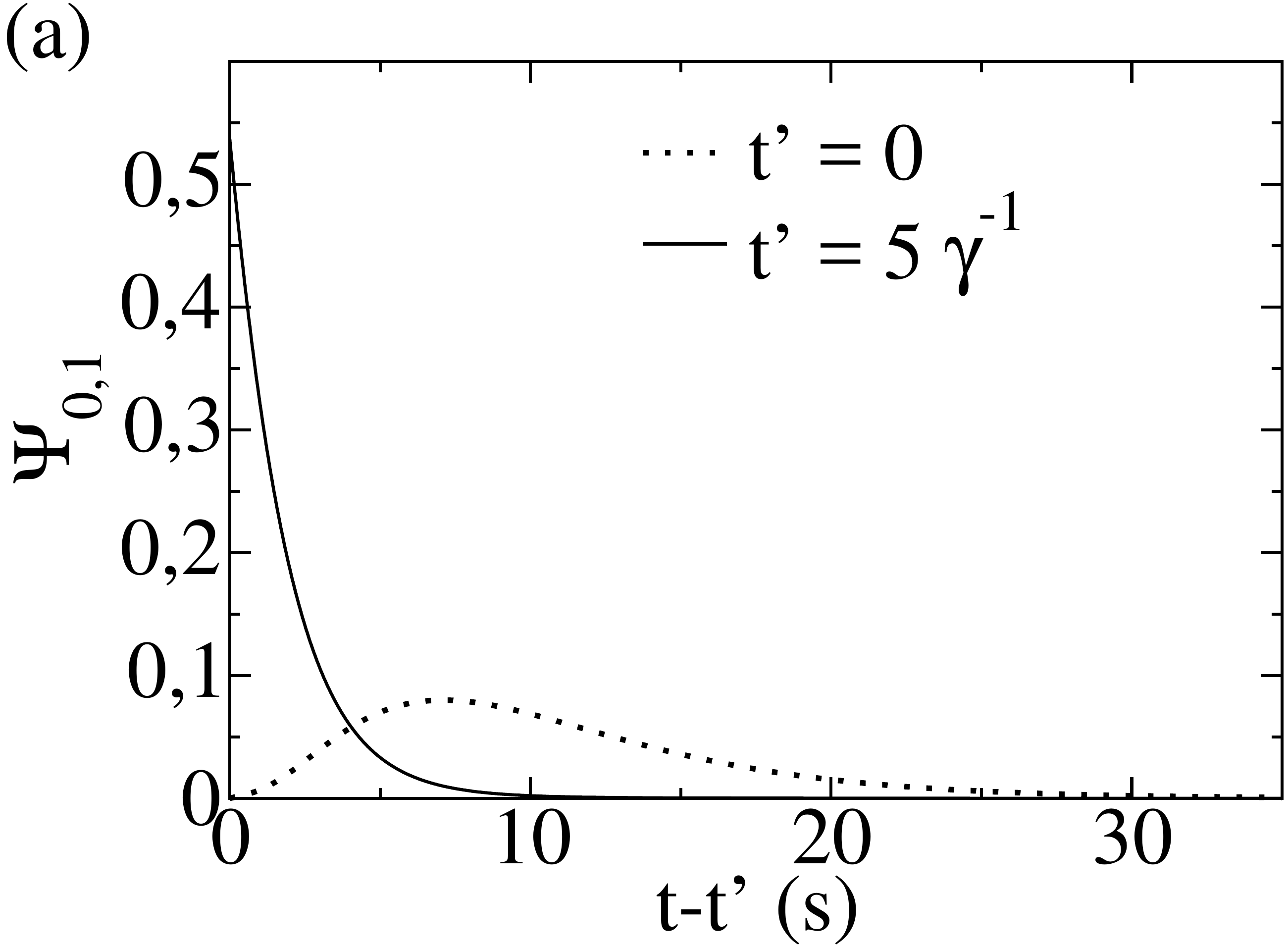}\includegraphics[width=.5\linewidth]{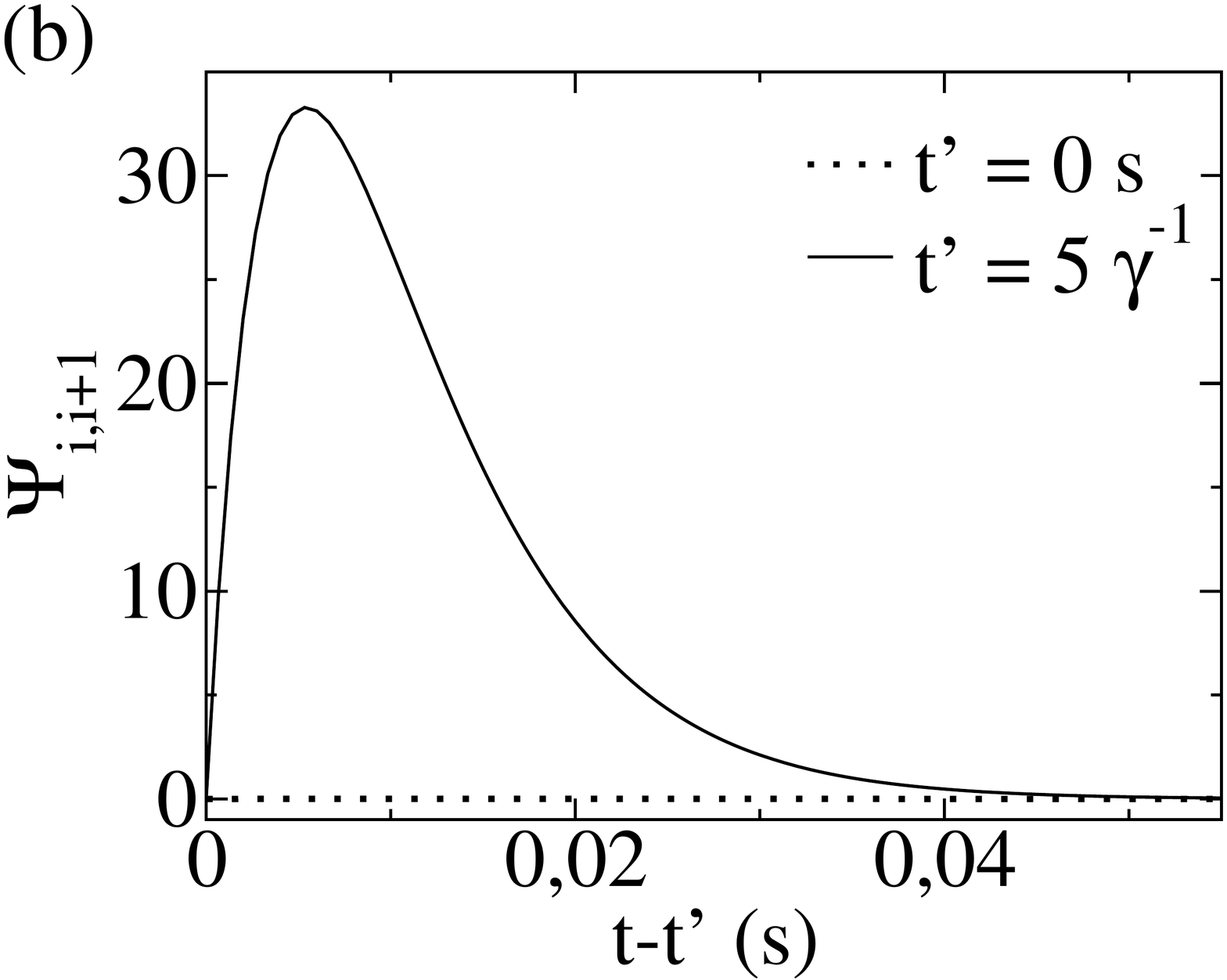}
\includegraphics[width=.5\linewidth]{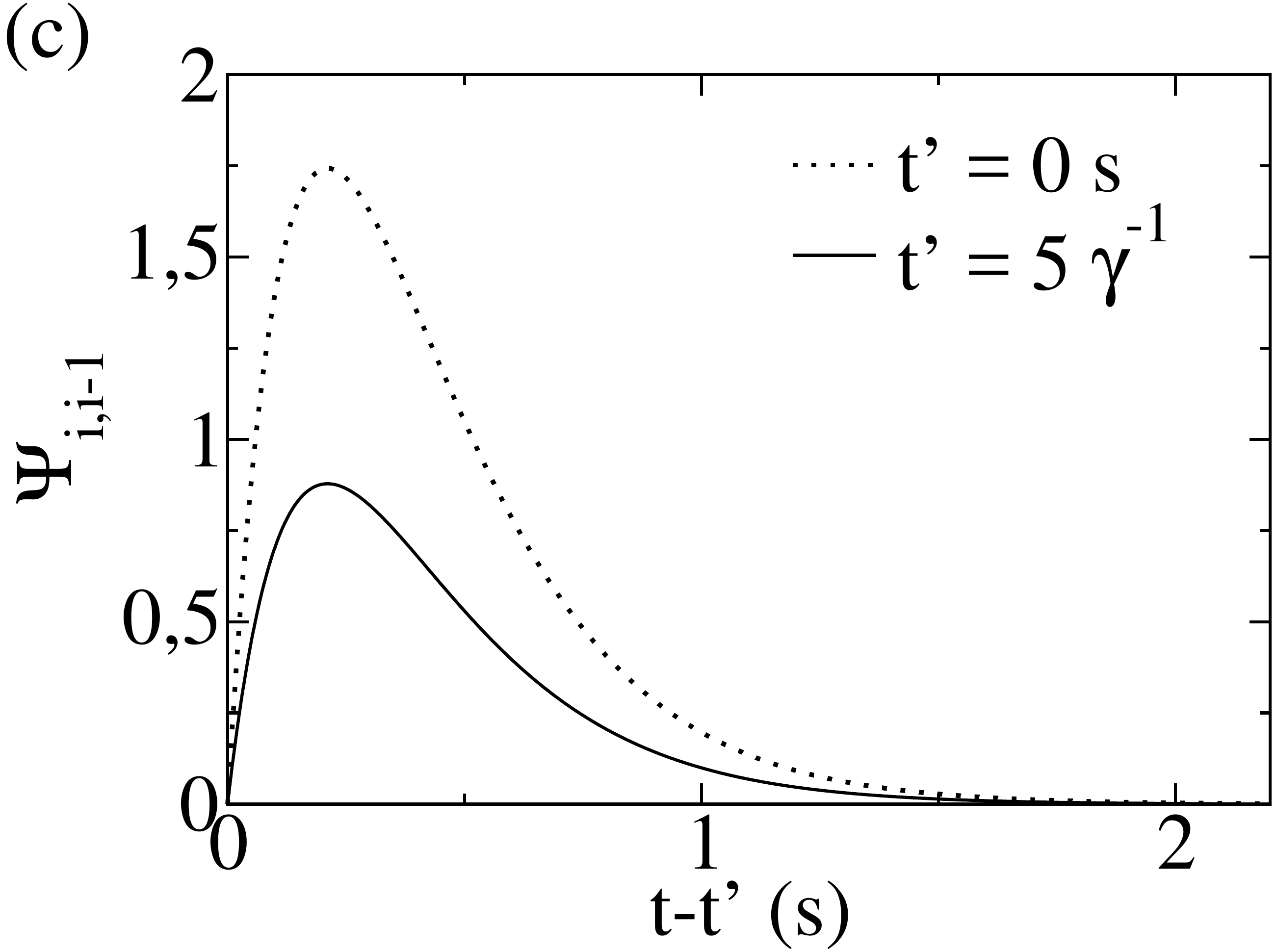}\includegraphics[width=.5\linewidth]{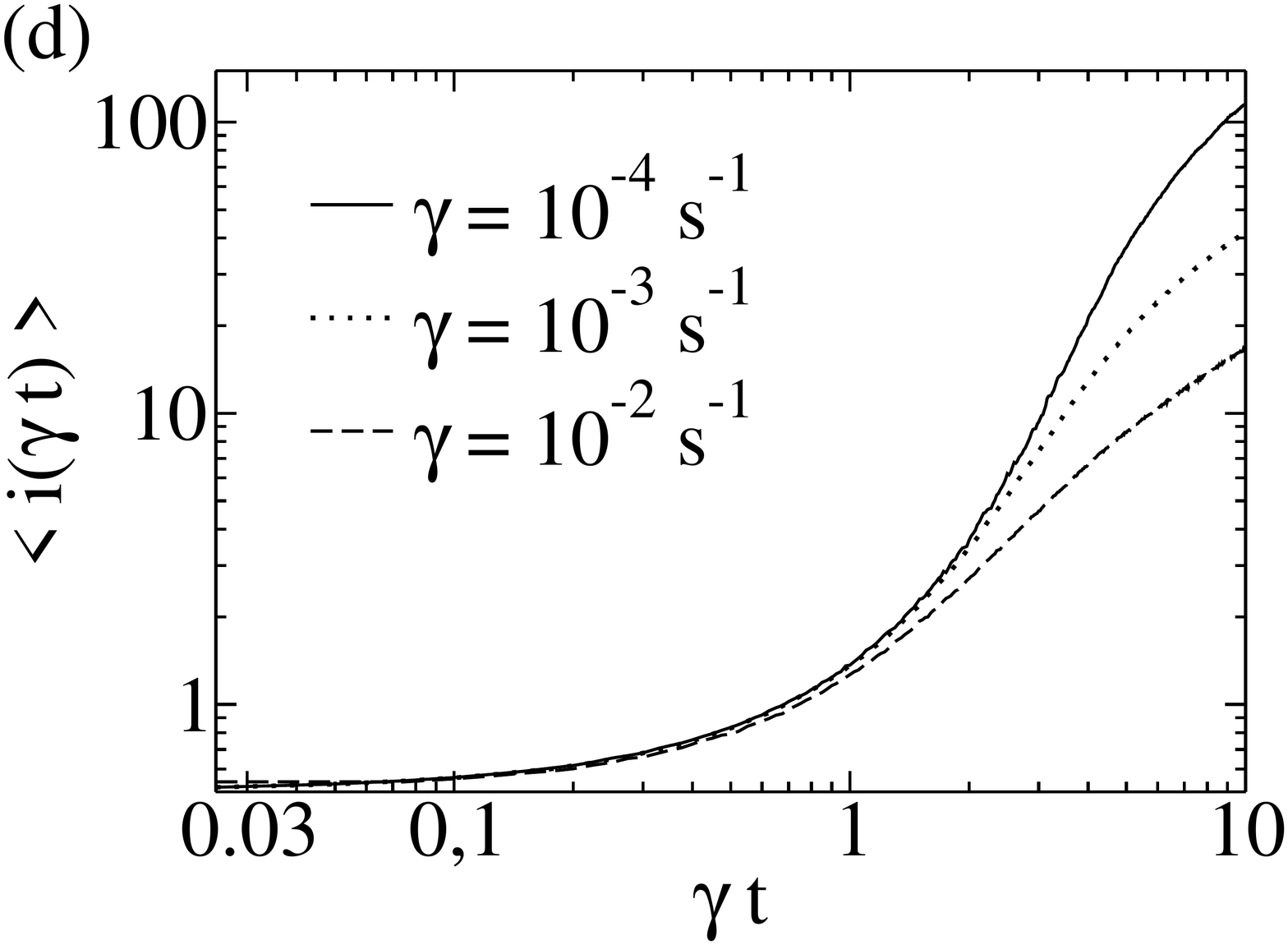}
\caption{Waiting-time distributions. (a) $\Psi_{0,1}$ with $\alpha_0$=0.6~s$^{-1}$, $\beta_0$=1.07~s$^{-1}$, $\delta_0$=0.25~s$^{-1}$, $\epsilon_0$=0.225~s$^{-1}$, $Z_0$=0.25. (b) $\Psi_{i,i+1}$ with $\alpha$=160~s$^{-1}$, $\beta$=211~s$^{-1}$, $\delta$=4.5~s$^{-1}$, $\epsilon$=5.0~s$^{-1}$. (c) $\Psi_{i,i-1}$ with the same parameter values as (b). (a--c) $\gamma$=0.01~s$^{-1}$. The parameter values of (a--c) are the standard parameter set, which are used if not mentioned otherwise. The dwell time in state 0 decreases owing to the slow transient, but is always in the range of seconds (a). Upon entering state $i$, the dependences of $\Psi_{i,i+1}$ and $\Psi_{i,i-1}$ on $t-t'$ with these large values of $\alpha$ and $\beta$ entail transitions either to $i+1$ very early or to $i-1$ later [see the abscissa range in (b) and (c)]. (d) Since $C_{i,i-1}$ is initially close to 1, the process lingers around  state 0 until $t\approx\gamma^{-1}$, as shown by the simulation results for the average state index $\langle$i($\gamma$t)$\rangle$. When $C_{i,i-1}$ approaches $\frac 12$ at $t>\gamma^{-1}$,  states further away from 0 with larger index $i$ are also reached.}
\label{fig:Psis}
\end{figure}

We specify the input for the generalized master equation of the non-Markovian system as
\begin{eqnarray}%\label{eq:Psipm1}
\Psi_{i,i+1}&=&\frac{\big(e^{-\alpha(t-t')}- e^{-\beta(t-t')}\big)\left(1-e^{-\gamma t}\right)}{\dfrac{\beta-\alpha}{\alpha\beta}+\dfrac{(\beta-\alpha)(\epsilon+\gamma)(\delta+\gamma)}{\epsilon\delta(\alpha+\gamma)(\beta+\gamma)}},\label{eq:Psiiiplus1}\\[12pt]
\Psi_{i,i-1}&=&\frac{\big(e^{-\delta(t-t')}- e^{-\epsilon(t-t')}\big)\left(1+e^{-\gamma t}\right)}{\dfrac{(\epsilon-\delta)(\alpha+\gamma)(\beta+\gamma)}{\alpha\beta(\epsilon+\gamma)(\delta+\gamma)}+\dfrac{\epsilon-\delta}{\epsilon\delta}} . \label{eq:Psiiiminus1}
\end{eqnarray}
The denominators in Eqs.~\eqref{eq:Psiiiplus1} and \eqref{eq:Psiiiminus1} arise from the normalization to $C_{i,i+1}(t')+C_{i,i-1}(t')$=1 [Eq.~\eqref{eq:Splitnorm} with $C_{i,i}=0$]. We use identical waiting-time distributions for all transitions $i\rightarrow i+1$ ($i>0$) and all transitions $i\rightarrow i-1$. The process defined by Eqs.~\eqref{eq:Psiiiplus1} and \eqref{eq:Psiiiminus1}  also has a strong bias toward 0 in the beginning, and is approximately symmetric in the limit $t\rightarrow\infty$. Figure~\ref{fig:Psis} illustrates the $\Psi_{i,i\pm1}$ and their development with increasing $t'$. The limit of $\gamma \ll \alpha, \beta,\epsilon, \delta$ illustrates the consequences of the transient. The $C_{i,k}$ relax from 0 to an asymptotic value with rate $\gamma$, but the dwell times of individual states $i$ ($i\neq0$) are essentially constant (Fig.~\ref{fig:Psis}).

There is only one transition away from 0. That changes the normalization to $C_{0,1}(t')=1$ and we cannot use the form of Eqs.~\eqref{eq:Psiiiplus1} and \eqref{eq:Psiiiminus1} for $\Psi_{0,1}$. We use instead
\begin{align}
&\Psi_{0,1} = \frac{\alpha_0\beta_0}{\beta_0-\alpha_0 Z_0}\bigg[\big(e^{-\alpha_0(t-t')}-Z_0 e^{-\beta_0(t-t')}\big)\label{eq:Psi01}\\
%&&\frac{\alpha_0\beta_0}{\beta_0-\alpha_0 Z_0}\left(\left(e^{-\alpha_0(t-t')}-Z_0 e^{-\beta_0(t-t')}\right)\right.-\nonumber\\
&-e^{-\gamma t}\frac{(1-Z_0)(\epsilon_0+\gamma)}{\delta_0-\epsilon_0}\left(\frac{\delta_0+\gamma}{\epsilon_0+\gamma}e^{-\delta_0(t-t')}- e^{-\epsilon_0(t-t')}\right)\bigg] \nonumber\\
&\qquad= g_{0,1}(t-t') + h_{0,1}(t-t') e^{-\gamma t} .\label{eq:gh0definition}
\end{align}
The transient here causes  a decrease of the dwell time in state 0 (see Fig.~\ref{fig:Psis}).

The first-passage-time probability density  $F_{0,N}(t)$ provides the probability of arrival for the first time in state $N$ in $(t,t+dt)$ when the process started in state $0$ at $t$=0. It is given by the probability flux out of the state range from 0 to $N-1$:
\begin{eqnarray} \label{eq:ithMom}
F_{0,N}(t) &=& - \frac{d}{d t} \sum_{k=0}^{N-1} P_{0,k}(t).
\end{eqnarray}
We denote its Laplace transform by $\tilde F_{0,N}(s)$. The moments of the first-passage-time distribution are given by~\cite{VanKampen:01}
\begin{equation}\label{eq:moments}
\langle t^n\rangle = \left.(-1)^n \frac{\partial^n}{\partial s^n}\tilde F_{0,N}(s)\right|_{s=0}.
\end{equation}
$F_{0,N}(t)$ can be determined by solving the master equations,  setting state 0 as the initial condition and considering the state $N$ as absorbing, i.e., $\Psi_{N,N-1}\left(t,t-t'\right)\equiv0$:
\begin{equation}\label{schemeabsorb}
0\stackrel[\Psi_{i,i-1}]{\Psi_{0,1}}{\rightleftarrows}1\stackrel[\Psi_{i,i-1}]{\Psi_{i,i+1}}{\rightleftarrows}2 \ \ldots \ N-1\stackrel[]{\Psi_{i,i+1}}{\rightarrow}N.
\end{equation}
%\begin{equation}\label{schemeabsorb}
%0\
%\begin{matrix}
  %\Psi_{0,1} \\
 % \rightleftarrows \\
 % \Psi_{i,i-1} \\
%\end{matrix}
%\ 1\
%\begin{matrix}
 % \Psi_{i,i+1} \\
 % \rightleftarrows \\
 % \Psi_{i,i-1} \\
%\end{matrix}
%\ 2\ \ldots\ N-1\
%\begin{matrix}
%  \Psi_{i,i+1} \\
%  \rightarrow \\
%  \  \\
%\end{matrix}
%\ N.
%\end{equation}
%\begin{equation*}
%0\stackrel[\Psi_{1,0}]{\Psi_{0,1}}{\rightleftarrows}1\stackrel[\Psi_{2,1}]{\Psi_{1,2}}{\rightleftarrows}2 \ \ldots \ N-1\stackrel[]{\Psi_{N,N+1}}{\rightarrow}N.
%\end{equation*}
$F_{0,N}(t)$  captures not only the first-passage time $0\rightarrow N$, but also, to a good approximation,  transitions starting at states with indices larger than 0 (and $<N$), since, owing to the initial bias, the process quickly moves into state 0 first and then slowly starts from there.

\begin{figure}
\centering
\includegraphics[width=.5\linewidth]{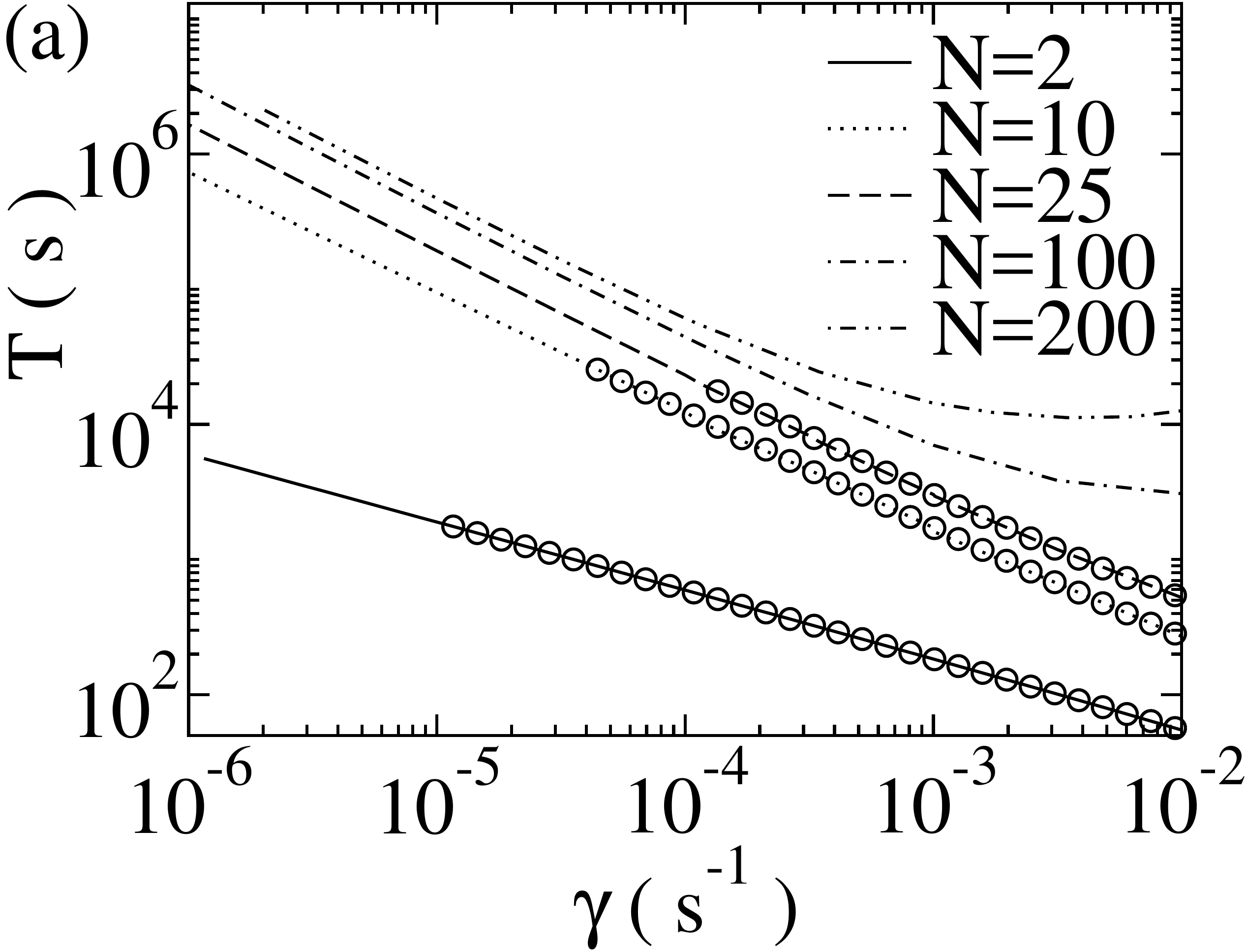}\includegraphics[width=.5\linewidth]{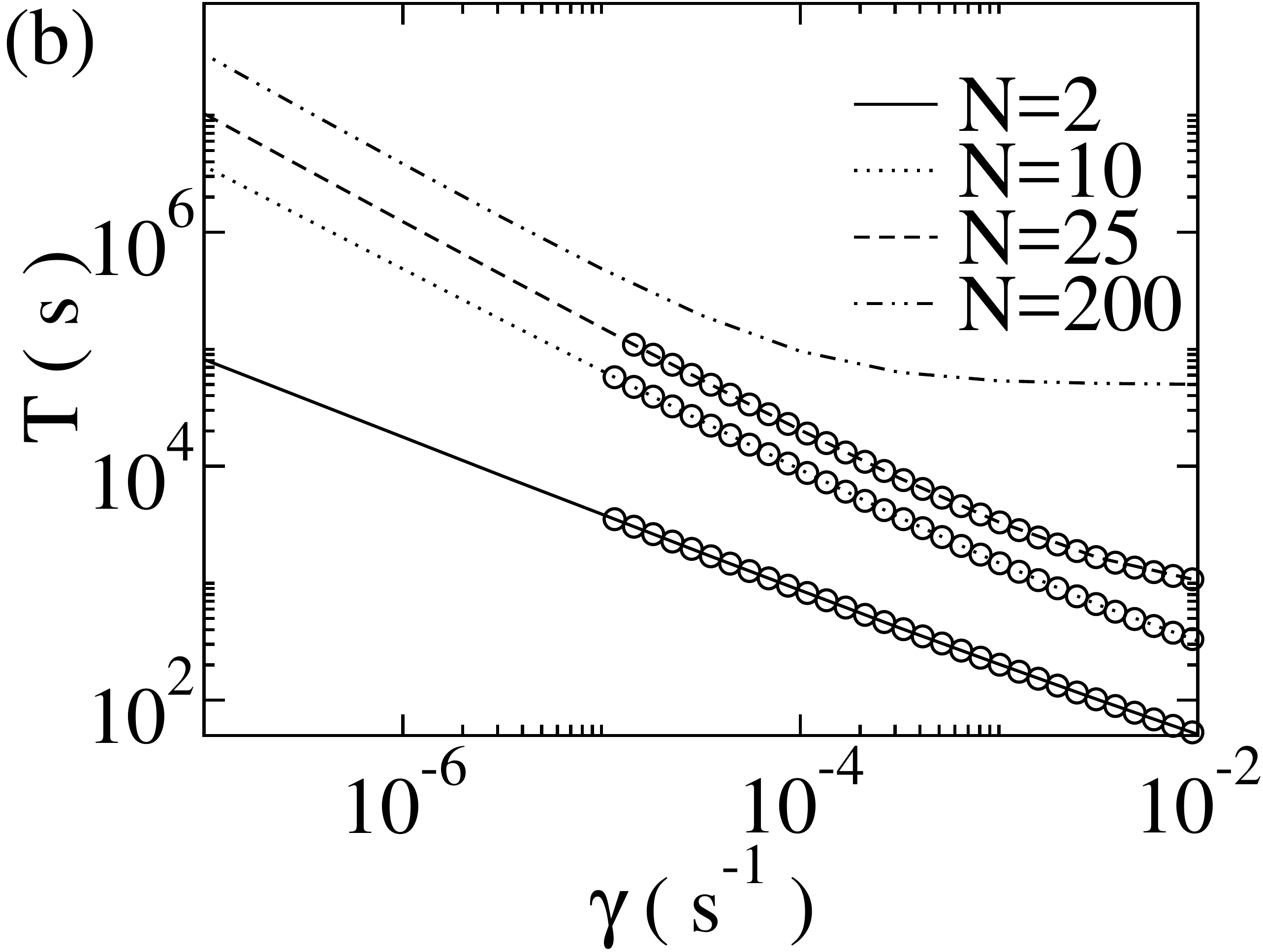}
\includegraphics[width=.48\linewidth]{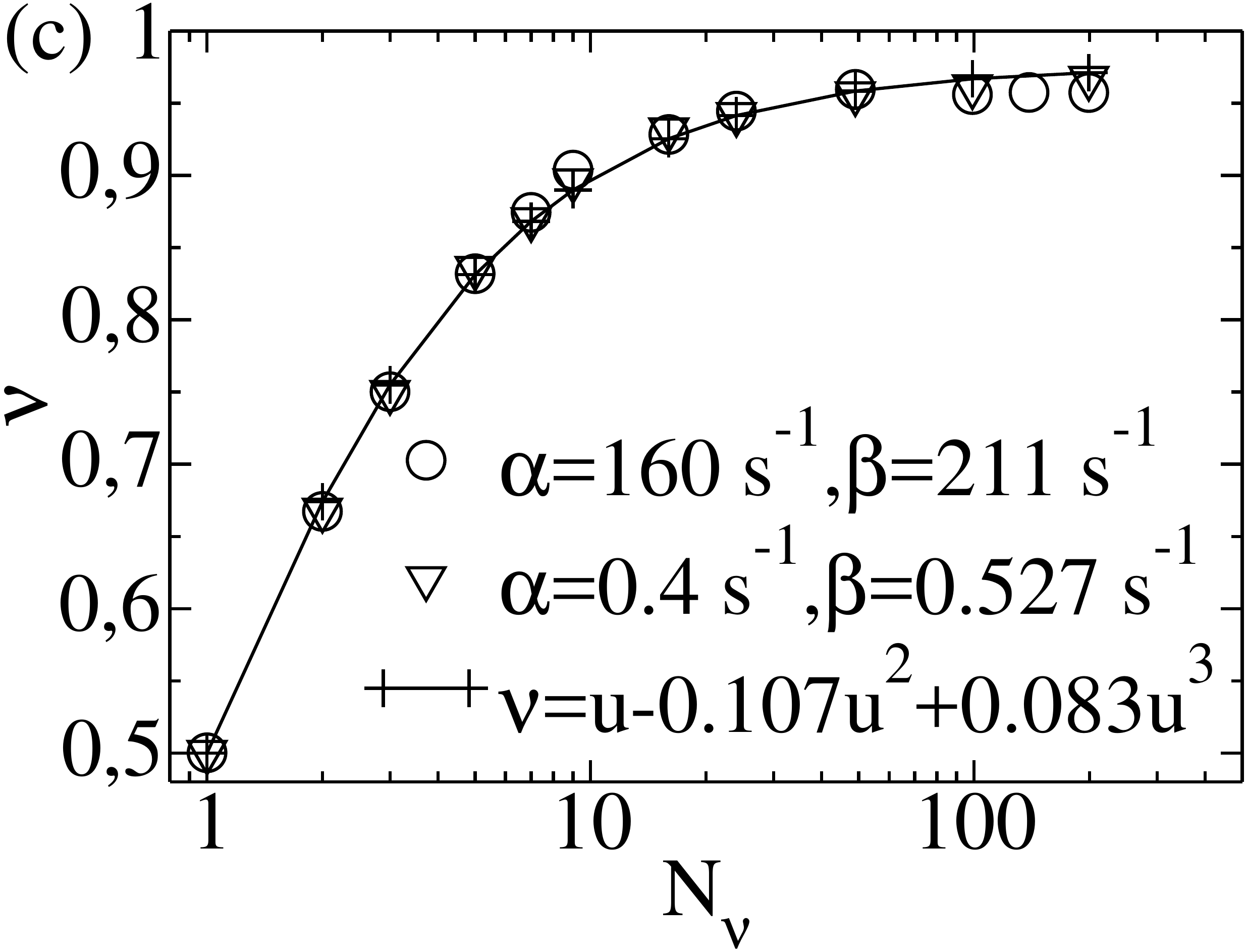}\hspace{0.2cm}\includegraphics[width=.48\linewidth]{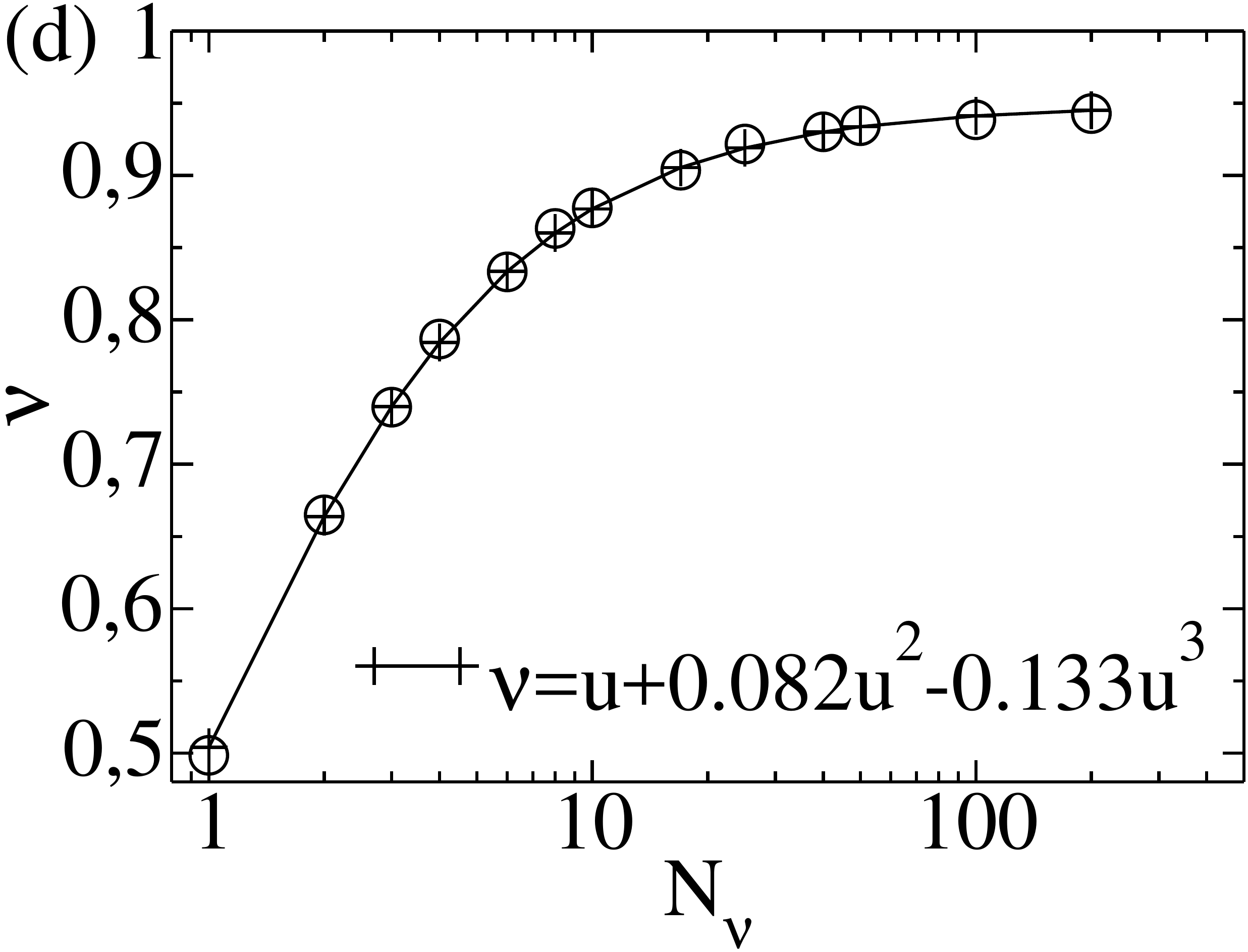}
\caption{The average first-passage time $T$ has a power-law dependence on the relaxation rate $\gamma$ of the initial transient of the form $\varpropto\gamma^{-\nu}$ for $\gamma\rightarrow 0$. (a, c) Results from  solution of the generalized master equation~\eqref{eq:Iinh} and Eqs.~\eqref{eq:Psiiiplus1}--\eqref{eq:Psi01}. (b, d) Results from solution of the asymptotically Markovian master equation~\eqref{eq:Marksolution} and Eq.~\eqref{eq:Mcase1} with $\lambda=0.4$~s$^{-1}$. $N_\nu$ is the number of edges with identical waiting-time distributions. $N_\nu$ is equal to $N-1$ in (c) and to $N$  in (d). The variable $u$ in (c) and (d) is defined as $u=N_\nu/(N_\nu+1)$. In (a) and (b), simulations (lines) are compared with analytical results ($\circ$).}
\label{fig:TavScaling}
\end{figure}

With the non-Markovian waiting-time distributions, the matrices $\tilde{G}(s)$ and $\tilde{H}(s)$ and the vector $\tilde{q}^i(s)$ of Eq.~\eqref{eq:Iinh} specific to this problem are
\begin{equation}\label{eq:matrixspecification}\begin{aligned}
  \tilde{G}(s)_{1,2} &= \tilde{g}_{0,1}(s),\\
  \tilde{G}(s)_{2i,2i-1} &= \tilde{G}(s)_{2i,2i+2}\\& =  \tilde{g}_{i,i-1}(s),\quad i=1,\ldots, N-2,\\
  \tilde{G}(s)_{2i+1,2i-1} &= \tilde{G}(s)_{2i+1,2i+2} \\&=  \tilde{g}_{i,i+1}(s),\quad i=1,\ldots, N-2,\\
  \tilde{G}(s)_{2N-2,2N-3}&=  \tilde{g}_{N-1,N-2}(s),\\
  \tilde{G}(s)_{2N-1,2N-3} &=  \tilde{g}_{N-1,N}(s),\\
  \tilde{H}(s)_{1,2} &= \tilde{h}_{0,1}(s),\\
  \tilde{H}(s)_{2i,2i-1} &= \tilde{H}(s)_{2i,2i+2} \\&=  \tilde{h}_{i,i-1}(s),\quad i=1,\ldots, N-2,\\
  \tilde{H}(s)_{2i+1,2i-1} &= \tilde{H}(s)_{2i+1,2i+2} \\&=  \tilde{h}_{i,i+1}(s),\quad i=1,\ldots, N-2,\\
%\end{eqnarray}
%\begin{eqnarray}
%  \tilde{G}(s)_{2N,2N-1} &=&  \tilde{g}_{N-1,N}(s)\nonumber
  \tilde{H}(s)_{2N-2,2N-3}&=  \tilde{h}_{N-1,N-2}(s),\\
  \tilde{H}(s)_{2N-1,2N-3} &=  \tilde{h}_{N-1,N}(s),\\
  \tilde{q}^i_{1}(s)&=\Psi_{0,1}(s).
\end{aligned}\end{equation}
All other entries are equal to 0. Specifically for the first-passage problem, the matrices $D$ and $E$ are
\begin{align*}
  E_{1,1}&=-E_{i,i\pm1}=-E_{1,2}=\lambda, \quad E_{i,i}=2\lambda, \\
  D_{1,1}&=D_{i,i}=-D_{i,i-1}=\lambda,\quad i=2,\ldots,N-1,
\end{align*}
with all other entries being 0. The vector $\tilde{r}$ is equal to $\delta_{1i}$, $i=1,\ldots,N-1$. The Laplace transforms of the first-passage-time distributions are
\begin{eqnarray}
  \tilde{F}_{0,N}(s)&=&\tilde{I}_{N-1,N}^0(s)\label{eq:FonsNM}
\end{eqnarray}
with Eq.~\eqref{eq:MasterIntDiff} and
\begin{eqnarray}
  \tilde{F}_{0,N}(s)&=&\lambda\big[\tilde{P}_{0,N-1}(s)-\tilde{P}_{0,N-1}(s+\gamma)\big]\label{eq:FonsM}
\end{eqnarray}
with Eq.~\eqref{eq:MasterEquation}. Figure~\ref{fig:TavScaling} (a) and (b) compare analytical results for the average first-passage time $T$ with the results of simulations. The agreement is very good,  thus confirming the solutions given by Eqs.~\eqref{eq:Iinh} and \eqref{eq:Marksolution}. This confirmation by simulations is important, since there is no method of solving difference equations that guarantees a complete solution.

\section{Scaling of the average first-passage time with the relaxation rate}\label{sec:scaling}

Figure~\ref{fig:TavScaling} shows results for the average first-passage time $T$ across four orders of magnitude of the relaxation rate $\gamma$. The results strongly suggest that $T$  grows according to a power law $\gamma^{-\nu}$ with decreasing $\gamma$, if $\gamma$ is sufficiently small. The exponent exhibits a simple dependence on the number $N_\nu$ of edges   with identical waiting-time distributions. That number is equal to $N$ for processes according to Eqs.~\eqref{eq:Mcase1} and  to $N-1$ for processes obeying  Eqs.~\eqref{eq:Psiiiplus1}-- \eqref{eq:Psi01}, since the $0\rightarrow1$ transition is different there. The exponent $\nu$ depends to leading order on $N_\nu$ like $N_\nu/(N_\nu+1)$. This applies to both the asymptotically Markovian and non-Markovian waiting-time distributions, and to both parameter sets of waiting-time distributions simulated in the non-Markovian case.

The exponent $\nu$ is equal to $\frac 12$ for $N_\nu=1$, as has previously been shown  analytically for $f_{i,i\pm 1}$ according to Eq.~\eqref{eq:Mcase1} (see~\cite{Calciumbook}, Chapter 5). The process is very unlikely to reach large $N$ with a bias toward 0, even if this is only small. Hence, the random walk ``waits'' until the transient is over and symmetry of the transition rates has been reached [see Fig.~\ref{fig:Psis}(d)], and then goes to $N$. This waiting contributes a time $\varpropto\gamma^{-1}$ to $T$, and $\nu$ approaches 1 for large $N_\nu$.

The average first-passage time for a symmetric random walk increases with $N$ like $N(N-1)$~\cite{Jouini2008}, i.e., it is very long for large $N$. We see a contribution of the transient to $T$ only if relaxation is slow enough for $\gamma^{-1}$ to be comparable to this long time. Consequently, $T$ is essentially independent of $\gamma$ for large $\gamma$ and large $N$ (see $N = 200$ in Fig.~\ref{fig:TavScaling}(a, b)).

\section{Resonant length}\label{sec:resonant}

\begin{figure}
  \centering
\includegraphics[width=.45\linewidth]{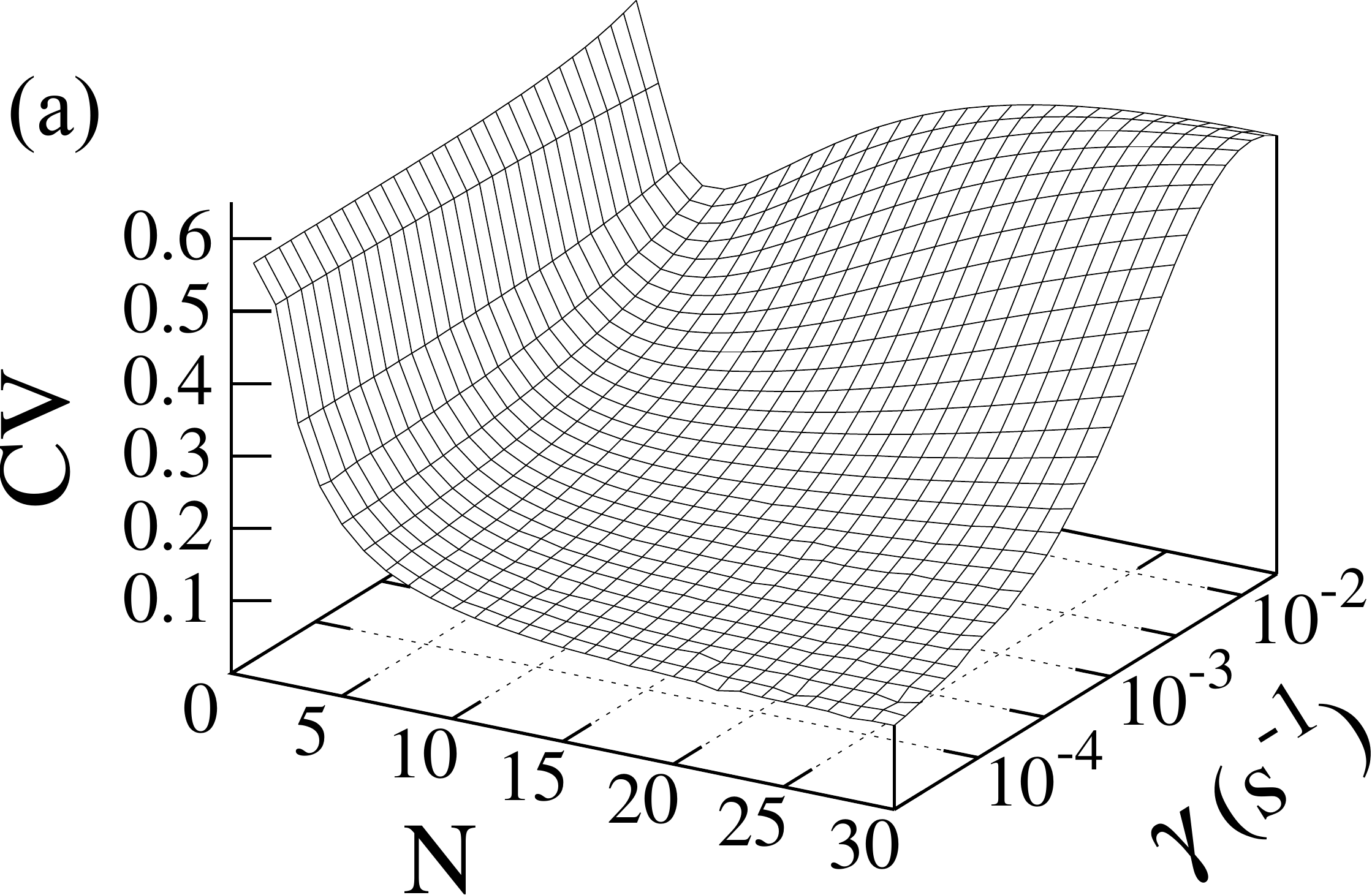}\hspace{0.2cm}\includegraphics[width=.45\linewidth]{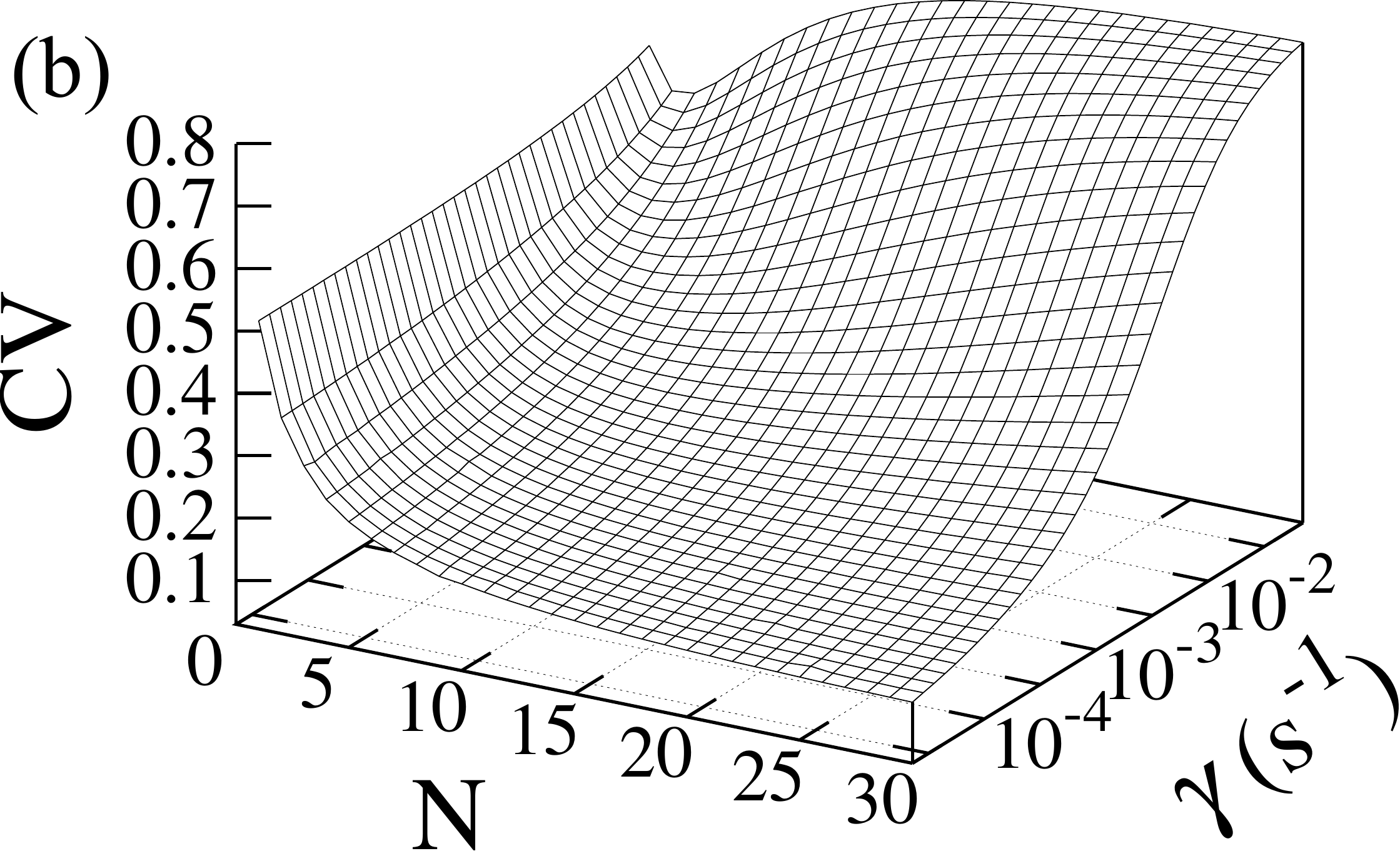}
\includegraphics[width=.5\linewidth]{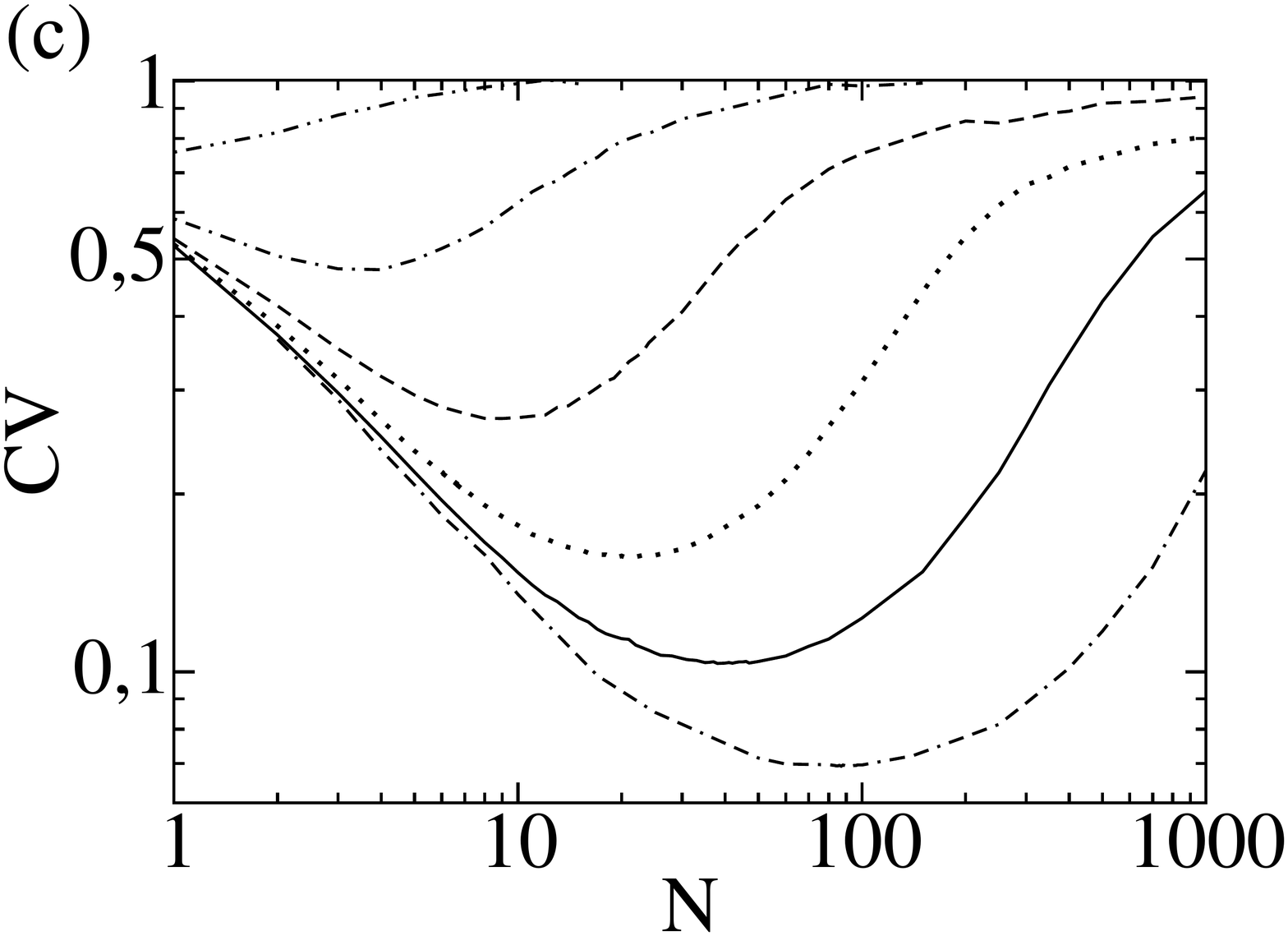}\includegraphics[width=.5\linewidth]{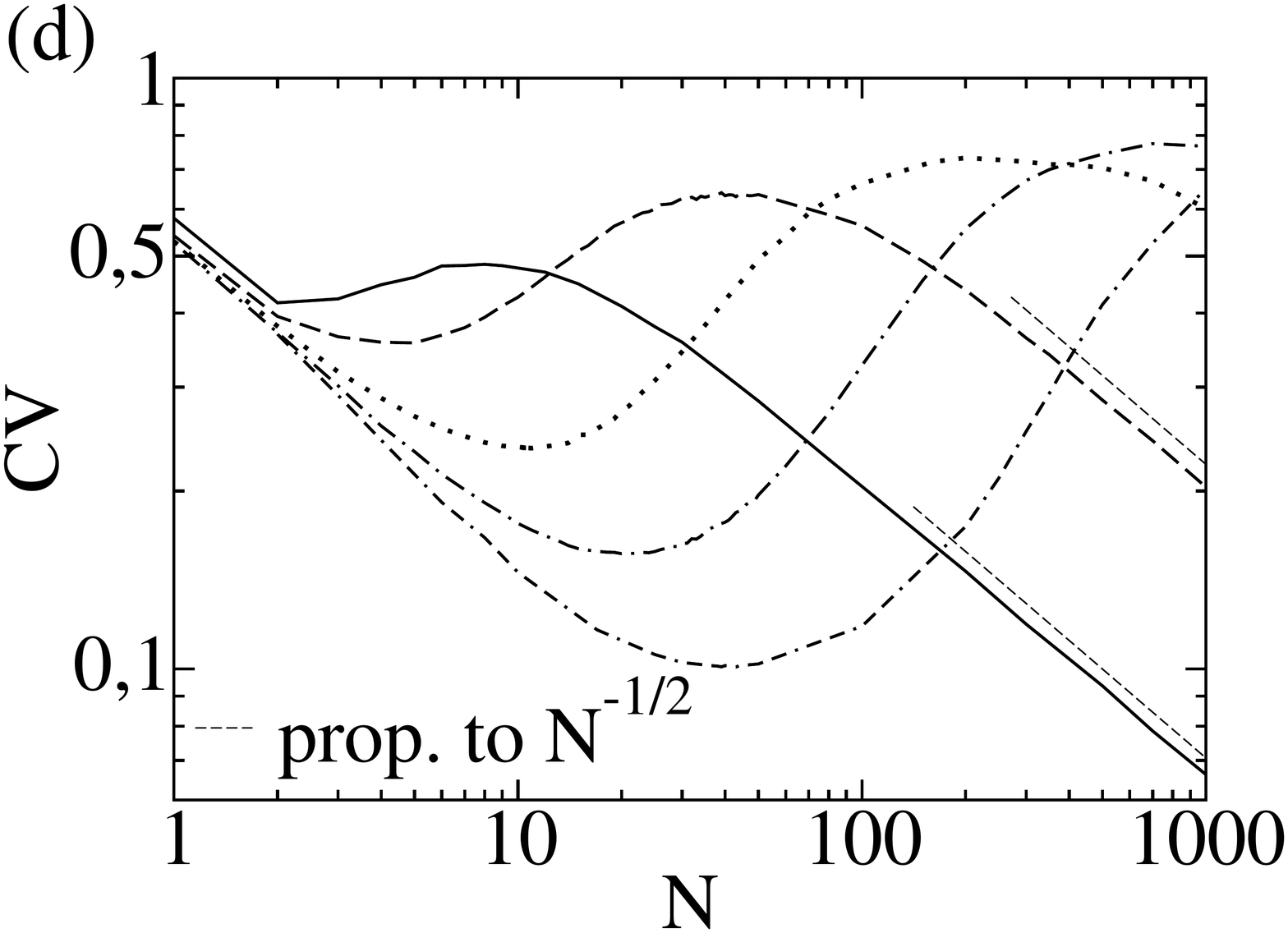}
\includegraphics[width=.5\linewidth]{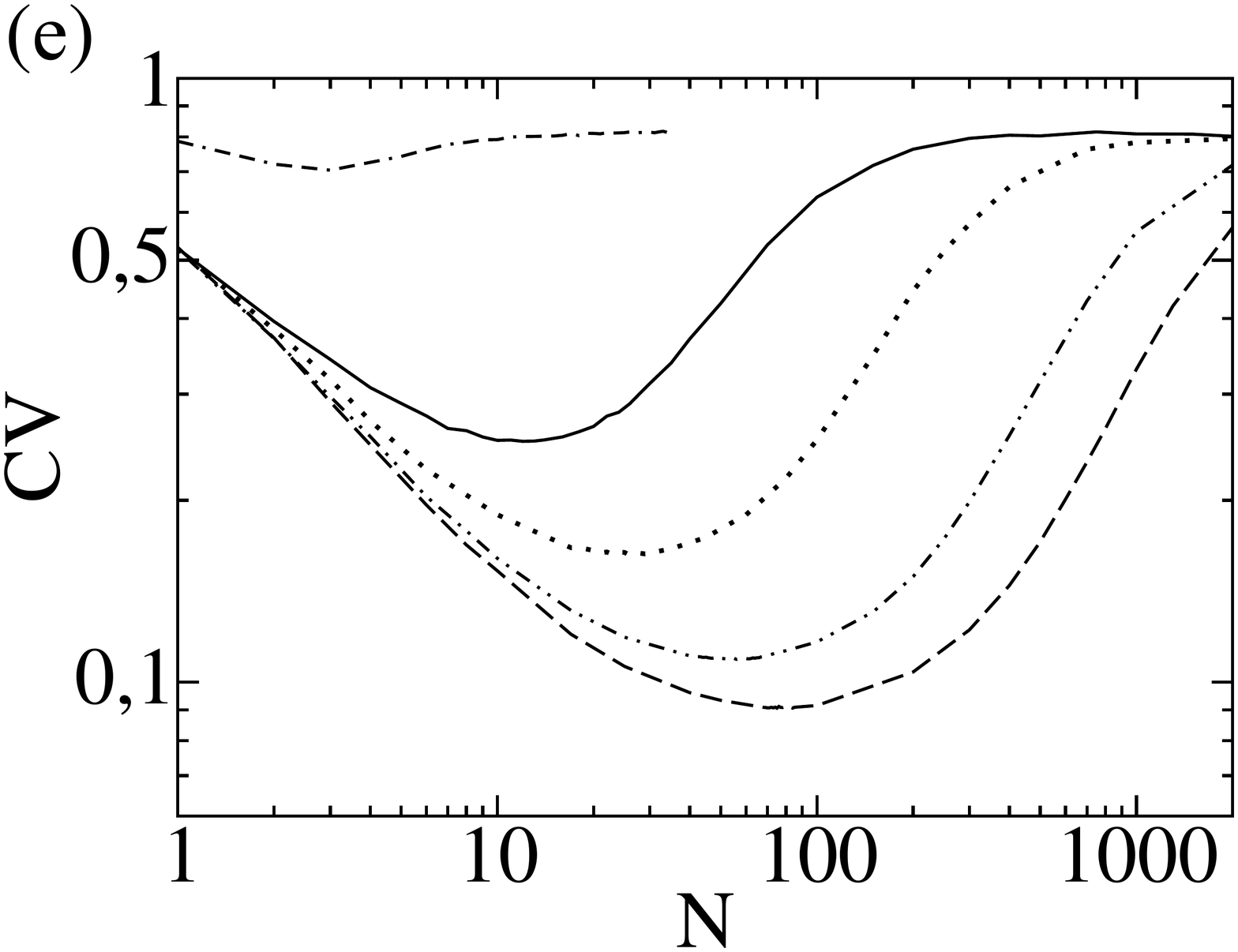}\includegraphics[width=.5\linewidth]{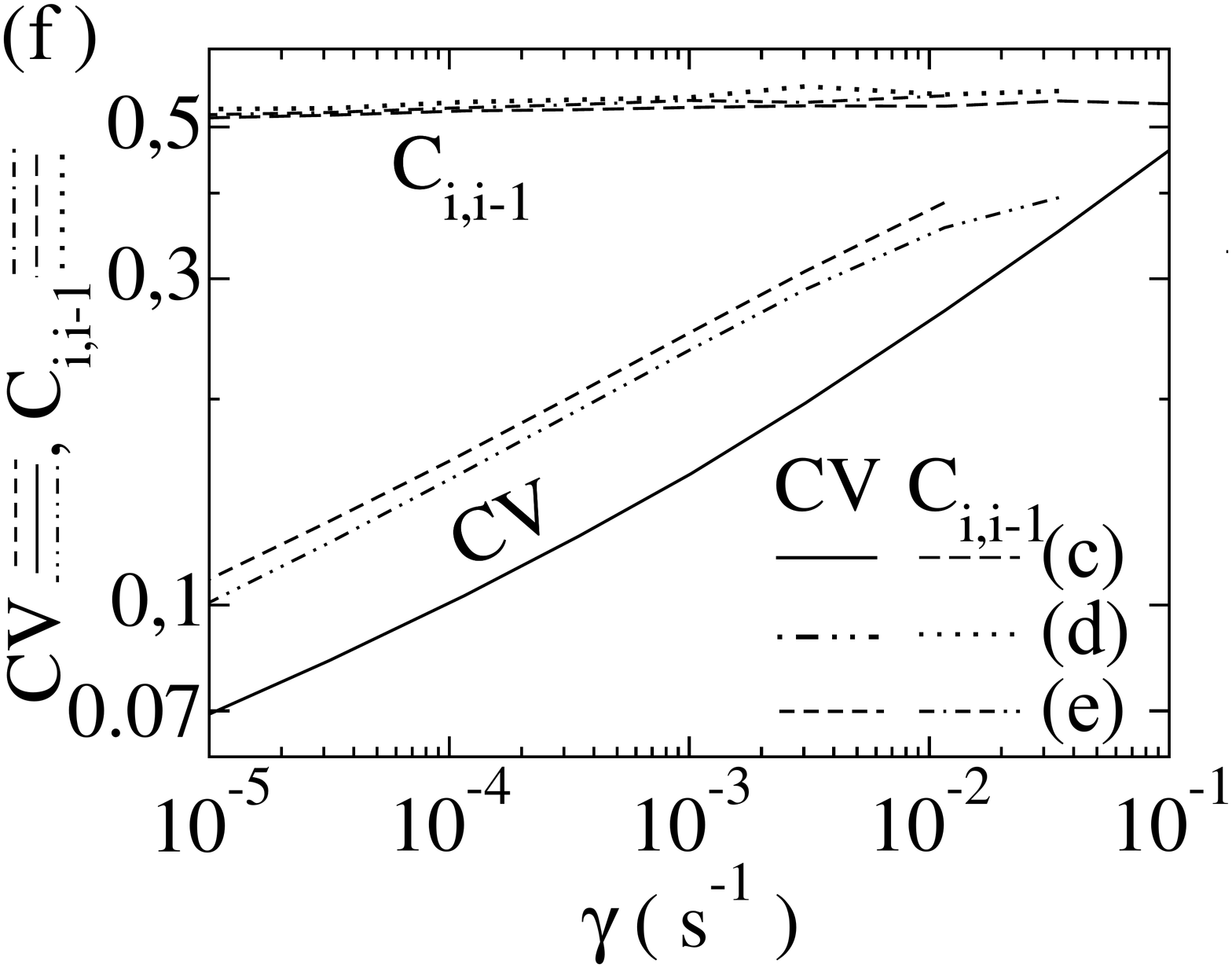}
\caption{The coefficient of variation CV shows a pronounced minimum in its dependence on the chain length $N$. (a) Analytical results using  the solution of the generalized master equation~\eqref{eq:Iinh} and Eqs.~\eqref{eq:Psiiiplus1}--\eqref{eq:Psi01} with $\alpha=0.4$~s$^{-1}$ and  $\beta=0.5275$~s$^{-1}$.  (b) Analytical results using the solution of the asymptotically Markovian master equation~\eqref{eq:Marksolution} and Eq.~\eqref{eq:Mcase1} with $\lambda=0.4$~s$^{-1}$.  (c, d) Simulations using Eqs.~\eqref{eq:Psiiiplus1}--\eqref{eq:Psi01}. In (c), from top to bottom, $\gamma=1$, $1.156\times 10^{-1}$, $1.156 \times10^{-2}$, $1.015\times 10^{-3}$, $1.156 \times10^{-4}$, and $10^{-5}$~s$^{-1}$; $C_{i,i-1} (t'=\infty)\gtrsim\frac 12$. In (d), $\alpha=0.4$~s$^{-1}$,  $\beta=0.5275$~s$^{-1}$, and, from top to bottom $\gamma=0.1$, $1.156 \times10^{-2}$, $1.015 \times10^{-3}$, $1.156\times 10^{-4}$, and $10^{-5}$~s$^{-1}$; $C_{i,i-1}(t'=\infty)\lesssim\frac 12$. (e) Simulations using Eqs.~\eqref{eq:Mcase1} with $\lambda=0.4$~s$^{-1}$, $C_{i,i-1}(t'=\infty)=\frac 12$, $\gamma=1$, $10^{-3}$, $10^{-4}$, $10^{-5}$, and $10^{-5.5}$~s$^{-1}$. (f) Properties of the minima of the CV as a function of the relaxation rate $\gamma$. The splitting probability $C_{i,i-1}(t'= T)$ at the minimum changes only slightly over four orders of magnitude of $\gamma$ [between 0.515 and 0.565 in (c), 0.531 and 0.564 in (d), and 0.516 and 0.555 in (e)].  The dependence of the minimal CV on $\gamma$ shown in (d) is well fitted by $\gamma^{0.181}$ in the asymptotically Markovian case (c), and by $\gamma^{0.186}$ for $\gamma < 2\times 10^{-3}$~s$^{-1}$ for the parameter values in (b), but deviates from a power law for the values in (a).}
\label{fig:CV}
\end{figure}
The coefficient of variation CV [= standard deviation (SD)/average $T$] of the first-passage time reflects the relative fluctuations. Its dependence on the chain length $N$ is illustrated in Fig.~\ref{fig:CV}.  CV increases monotonically with increasing $\gamma$. Its dependence on $N$ is not monotonic. We find a pronounced minimum of CV($N$) for small values of $\gamma$, where the minimal value is up to one order of magnitude smaller than  CVs with $N=1$ and with large $N$. Both simulations and analytical results show this behavior.

How can we get a heuristic understanding of the decrease in CV with decreasing $\gamma$ and initially with increasing $N$? The lingering close to state 0 shown in Fig.~\ref{fig:Psis}(d) means that states with index $i$ larger than 1 are not reached before a time $t\approx\gamma^{-1}$  with almost certainty. This initial part of the process contributes to $T$ but little to SD. Hence, its growth with decreasing $\gamma$ and (initially) increasing $N$ decreases CV. Additionally, the standard deviation is determined by the rates and splitting probabilities at the end of the transient, which are more favorable for reaching $N$ than the splitting probabilities at the beginning. It is therefore less affected  than the average by the $\gamma$ values. This also causes a decrease in CV with decreasing $\gamma$.

\begin{figure}
  \centering
\includegraphics[width=.48\linewidth]{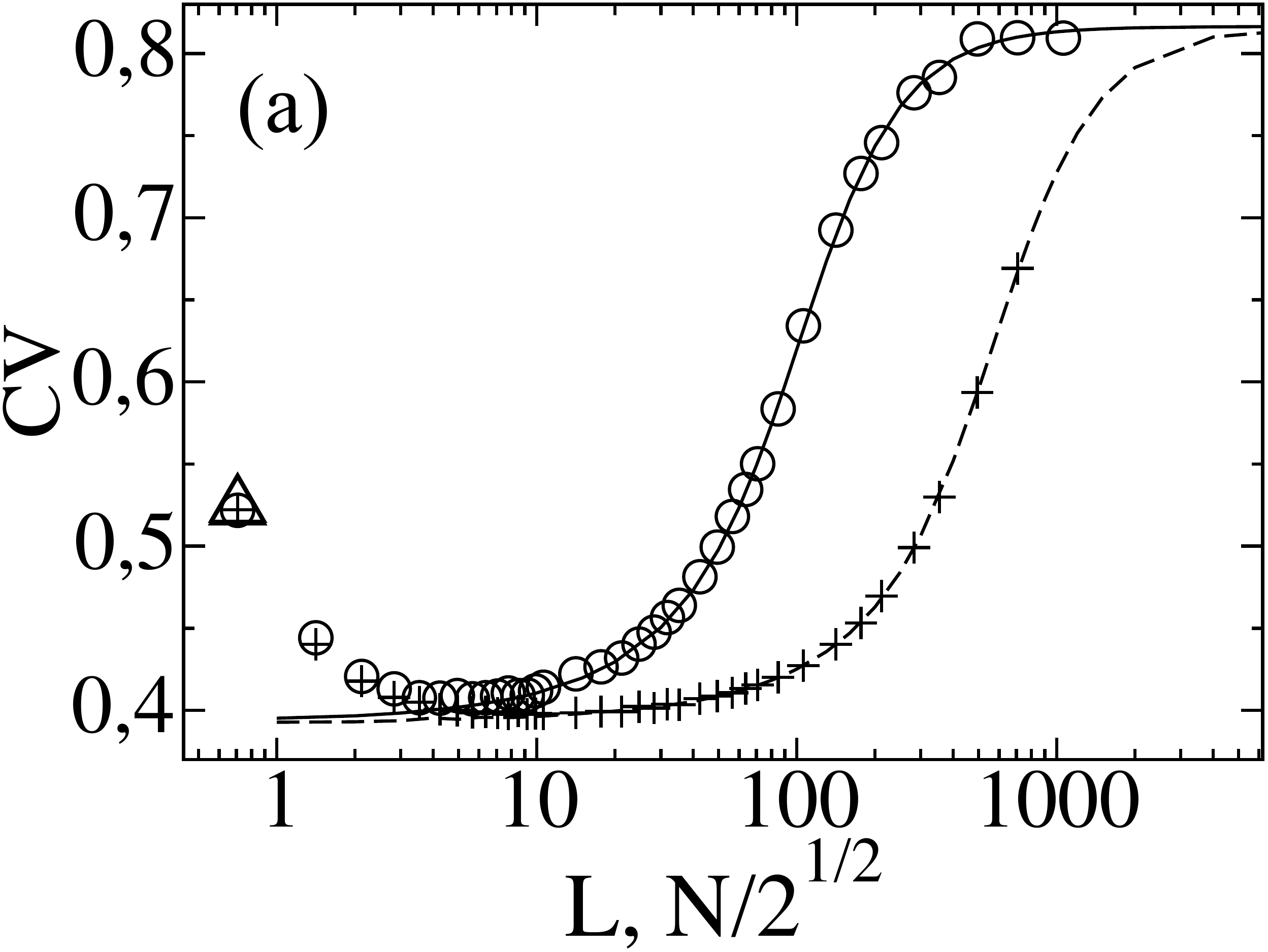}\hspace{0.1cm}\includegraphics[width=.5\linewidth]{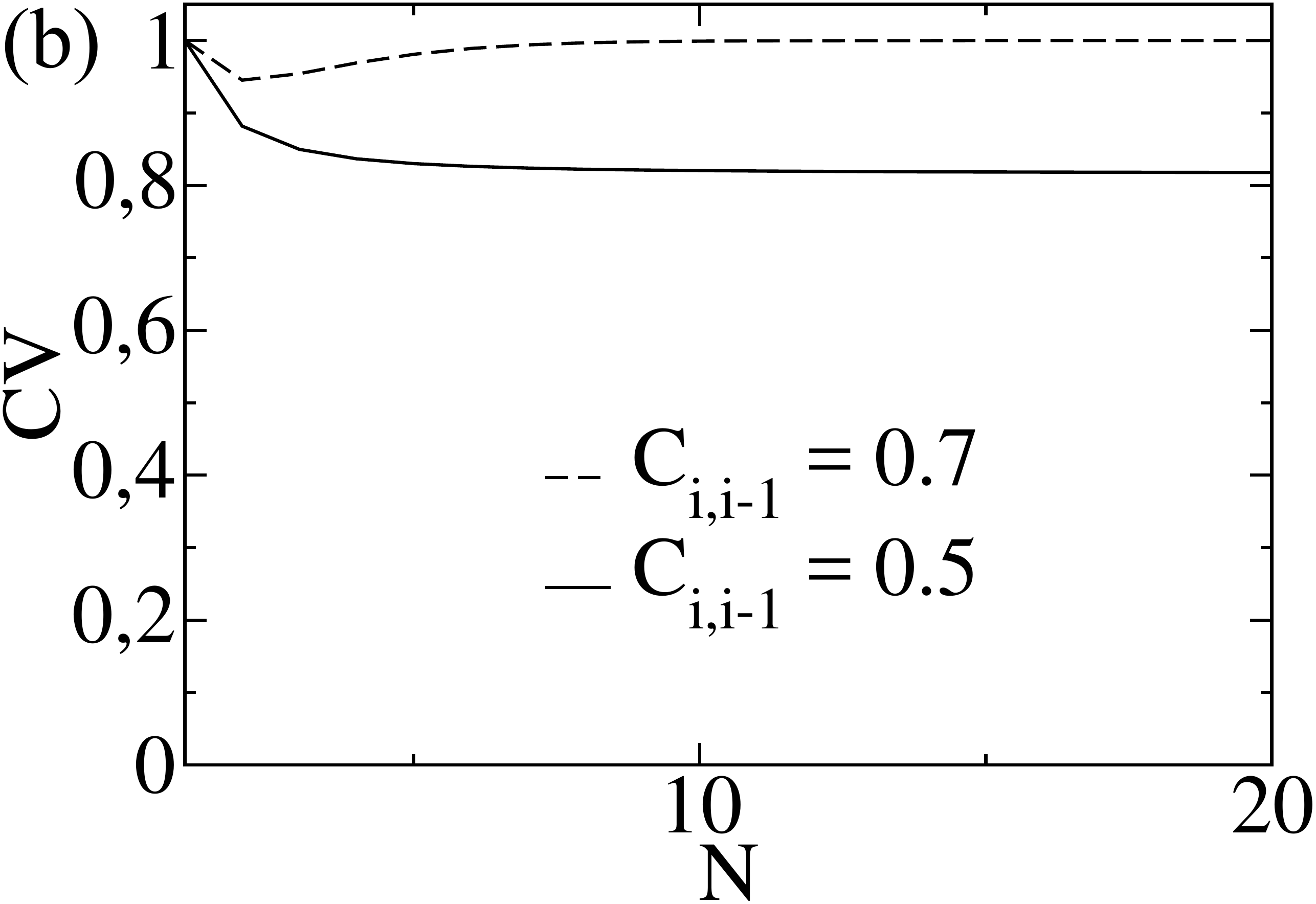}
\caption{(a) The coefficient of variation CV shows a shallow minimum in its dependence on the chain length $N$ with processes that are discrete and always symmetric (symbols), but not for the continuous symmetric case (lines), which is the solution of the Fokker--Planck equation~\eqref{eq:Contcase}. The symbols + and $\circ$ show results from simulations using Eq.~\eqref{eq:Mcase2}.  The calculations used $\lambda=0.4$~s$^{-1}$ (all), and $\gamma=10^{-4}$~s$^{-1}$ ($\circ$, full line),  $\gamma=3.6\times10^{-6}$~s$^{-1}$ (+, dashed line). The value of the CV at the minimum is essentially unaffected by the change in the value of $\gamma$. The $\bigtriangleup$ marks the  analytical result with $N=1$ for both discrete cases, which are indistinguishable at the resolution of the plot and are in agreement with the simulations. The continuous large-$L$ and discrete large-$N$ results agree very well. (b)  CV with constant transition rates calculated using the equations of Jouini and Dallery~\cite{Jouini2008}. CV $\geq\sqrt{2/3}$ holds for $C_{i,i-1}\geq\frac 12$.}
\label{fig:CVsym}
\end{figure}

To gain more insight into the $N$ dependence of  CV, we look at processes with constant transition rates and a symmetric process with transient rates but constant splitting probabilities $C_{i,i-1}=C_{i,i+1}=\frac 12$ [see Eq.~\eqref{eq:Mcase2}]. The permanently symmetric process exhibits only a very shallow minimum, and the minimal value of  CV appears to be independent of $\gamma$ [Fig.~\ref{fig:CVsym}(a): discrete case]. Processes with constant transition rates exhibit  decreasing CV with decreasing $C_{i,i-1}$~\cite{Jouini2008} [Fig.~\ref{fig:CVsym}(b)]. However, this decrease is minor and does not explain the data in Fig.~\ref{fig:CV} in the range $C_{i,i-1}\geq\frac 12$ covered by the transient.

The comparison shows that the dynamics of the splitting probabilities provide  the major part of the reduction in  CV  with increasing $N$ in Fig.~\ref{fig:CV}. The splitting probability $C_{i,i-1}$ relaxes from 1 to its asymptotic value set by the parameters of the process. The larger the value of $N$, the later is the absorbing state  reached and the smaller is the value of $C_{i,i-1}$ when it is reached. As long as $C_{i,i-1}(t'=T)$ decreases sufficiently rapidly with increasing $N$, so does  CV. At values of $N$ such that $\gamma T\gtrsim2$, the decrease in $C_{i,i-1}(t'=T)$ is negligible, and  CV starts to rise again with increasing $N$ toward its large-$N$ value.

These considerations suggests that the $N$ with minimal CV could be fixed by a specific value of the splitting probability. This value of $C_{i,i-1}$ cannot be smaller than $\frac 12$, since we would expect a monotonically decreasing CV in that regime [the $N^{-\frac 12}$ regime in Fig.~\ref{fig:CV}(d)]. Since this should hold also for minima with $N\gg1$, this value of $C_{i,i-1}$  also cannot be much larger than $\frac 12$, since $T$ would then diverge. This is confirmed by the results shown in Fig.~\ref{fig:CV}(f). The splitting probability $C_{i,i-1}(t'=T)$ at the minimum changes by less than 8\% over four orders of magnitude of $\gamma$, and is slightly larger than $\frac 12$. Hence,  CV starts to rise again when the length $N$ is so large that the average first-passage time is long enough for  $C_{i,i-1}(t'=T)$ to approach the symmetric limit~\footnote{The minima of the CV are less correlated with a specific value of the product $\gamma T$, since $\gamma T$ at the minima changes by about 20\% in Fig.~\ref{fig:CV}(c) and about 80\% in Fig.~\ref{fig:CV}(d) and (e) across the $\gamma$-range.}.

The value of CV($N=\infty$)  depends on $\gamma$ in the case with non-Markovian waiting-time distributions. It is 1 for large $\gamma$ values, since the asymptotic splitting probability $C_{i,i-1}(t'=\infty)$ is larger than $\frac 12$ [Fig.~\ref{fig:CV}(c); see also Fig.~\ref{fig:CVsym}(b)]. CV($N=\infty$) is approximately equal to $\sqrt{2/3}$ for small values of $\gamma$, since $C_{i,i-1}(t'=\infty)\approx\frac 12$ applies in that case [Fig.~\ref{fig:CV}(c)]. CV($N=\infty$) is exactly $\sqrt{2/3}$ with asymptotically Markovian waiting-time distributions [Fig.~\ref{fig:CV}(e)].

The parameter values in Fig.~\ref{fig:CV}(d) entail $C_{i,i-1}(t'=\infty)\lesssim\frac 12$. The process has a bias toward $N$ and we see the well-known behavior CV($N$) $\propto N^{-\frac 12}$ for large $N$. The onset of the $N^{-\frac 12}$ behavior moves to smaller $N$ with increasing values of $\gamma$ until finally the minimum of the CV is lost. We found minima if $\gamma^{-1}\gtrsim$ (four times the average state dwell time), but we have not determined a precise critical value.

The transition from the CV of $\Psi_{0,1}$ to CV($N=\infty$) for large values of $\gamma$ is monotonic in Fig.~\ref{fig:CV}(c) for the asymptotically non-Markovian case. The asymptotically Markovian system exhibits a minimum even for large $\gamma$ values [$\gamma\approx\lambda$: Fig.~\ref{fig:CV}(e)]. However, it is comparably shallow, and is in line with the results of Jouini and Dallery for systems without transient~\cite{Jouini2008} [Fig.~\ref{fig:CVsym}(b)].

\section{Discussion and Conclusion}
\subsection{(Generalized) master equation with exponential time dependence of rates and waiting-time distributions}
In general, stochastic processes occur in changing environments or may be subjected to control. The corresponding mathematical description is provided by (generalized) master equations with time-dependent coefficients. The analytical solution of these equations is complicated even in the  simplest case
%(asymptotically Markovian, $N=1$) leads to expressions for the moments that involve generalized hypergeometric functions, which are notoriously difficult to evaluate
[see Eq.~\eqref{eq:SecondMom}]~\cite{falcke2010_EPJSTa}. The Laplace-transform approach is not applicable in general, but the exponential time dependences of  the master equation~\eqref{eq:MasterEquation} or the  generalized master equation~\eqref{eq:MasterIntDiff} with \eqref{eq:probfluxes} do allow the transform to be performed on them,  leading to difference equations in Laplace space. The solutions of these equations are given by Eqs.~\eqref{eq:Iinh} and \eqref{eq:Marksolution}, resp. %, and provide the Laplace transforms of the probability fluxes, state probabilities, and first-passage-time distributions.
To the best of our knowledge, this is a novel (and efficient) method of solving the generalized master equation with this type of time dependence in the kernel, or the master equation with this type of time-dependent rates.

Equations~\eqref{eq:Iinh} and \eqref{eq:Marksolution} also apply in the case of state schemes more complicated than that in \eqref{scheme}. It is only necessary to make appropriate changes to the  definitions of $\tilde{H}(s)$ and $\tilde{G}(s)$ or of $D$ and $E$. Hence, we can  use the method presented here to investigate the dynamics of complicated networks in a transient. This is of interest for studies of ion-channel dynamics~\cite{falcke2005c_Sneyd_comparison,Falcke2009_GinBJ}, transport in more than one spatial dimension, gene regulatory networks, and many other applications. Many quantities of interest in these investigations are moments of first-passage type and thus can be calculated from the derivatives of the Laplace transforms. Some of these calculations might require knowledge of the probability time dependence $P_{i,j}(t)$, i.e., the residues of $\tilde{P}_{i,j}(s)$. In many cases, determination of these residues will be only  slightly more complicated than for the system without transient, since all terms arising from the transient involve the same factors, but with shifted argument, together with the matrix $\tilde{H}(s)$. In general, if we know the set of residues $\{S_0\}$ of the Laplace transform of the system without transient and the matrix $\tilde{H}(s)$, we can immediately write down the residues of the Laplace transform of the system with transient by shifting $\{S_0\}$ by integer multiples of $\gamma$. This allows  easy generalization of many results and might be of particular interest for renewal theory~\cite{Coxrenewal,VanKampen:01}.

\subsection{The average first-passage time}
The average first-passage time decreases with increasing relaxation rate according  to a power law, if the transient is sufficiently slower than the time scale of the individual steps (Fig.~\ref{fig:TavScaling}). The exponent depends on the chain length to leading order like $-N/(N+1)$. Remarkably, this behavior has been found for substantially different parameters in both the asymptotically non-Markovian and asymptotically Markovian cases, and therefore appears to be rather universal.

\subsection{Resonant length}
Transients can substantially reduce the coefficient of variation. We have found a monotonic decrease of  CV with decreasing $\gamma$ when the relaxation rate is slow compared with state transition rates.  CV exhibits a minimum in its dependence on the chain length $N$ (Fig.~\ref{fig:CV}) if the transient is slow compared with state dynamics. This minimum exists for both asymptotically Markovian and non-Markovian discrete systems.  At a time-scale separation between the state transition rates and the transient of about $10^6$, the CV at the minimum is reduced by about one order of magnitude compared with  CV($N=1$).

Exploiting our results for control purposes, we can use a transient if more precise arrival timing at the outcome of a process is desired. Applying a transient that relaxes a bias toward 0 reduces CV. In the context of  charge carrier transport, such a transient could be realized by a time-dependent electric field. If we include the rising part of CV($N$) beyond the minimum in our consideration and consider all CVs smaller than CV($N=1$) as reduced, we may still  see a reduction of  CV even if the average first-passage time is an order of magnitude longer than $\gamma^{-1}$. Hence, the initial transient may have surprisingly long-term consequences and is therefore a rather robust method for CV reduction. Additionally, or if a transient is given, the state and step number can be used for optimizing the precision of arrival time. The optimum may also be at smaller lengths than that of the non-optimized process; i.e., optimization of precision may even lead to acceleration.

\subsection{How can negative feedback robustly reduce noise in timing?}
Most studies investigating the effect of negative feedback have focused on amplitude noise---we are looking at noise in timing.  Timing noise substantially reduces information transmission in communication systems. Variability in  timing and protein copy numbers in gene expression or cell differentiation causes cell variability~\cite{nachman2012,elowitz2017a}. Therefore, many  studies have investigated the role of noise in gene regulatory networks and have found that immediate negative feedback is not suitable for reducing timing noise~\cite{McArthur2016,Hinczewski2016,Ghusinga2017}. This agrees with our results for large $\gamma$ values.

We have shown that slow recovery from negative feedback is a robust means of noise reduction. The recovery process corresponds to the slow transient in our present study. Negative feedback terminating the excited state is a constitutive element of many systems generating sequences of pulses or spikes, such as oscillating chemical reactions~\cite{engel2015,falcke94}, the electrical membrane potential of neurons~\cite{Kandel91}, the sinoatrial node controlling the heart beat~\cite{leonhardt2016}, the tumor suppressor protein p53~\cite{Vousden2007,falcke2017_moenke}, Ca$^{2+}$ spiking in eukaryotic cells~\cite{Falcke04,Falcke2008,falcke2014_SciSig}, cAMP concentration spikes in \emph{Dictyostelium discoideum} cells and other cellular signaling systems~\cite{Goldbeter,Falcke98}.

In particular, our results apply to noisy excitable spike generators in cells, where single molecular events or sequences of synaptic inputs form a discrete chain of states toward the threshold. Beside examples from membrane potential spike generation~\cite{benda2017}, Ca$^{2+}$ spiking and p53 pulse sequences have been shown to be noise-driven excitable systems~\cite{falcke2017_moenke,falcke2014_SciSig,Falcke2008,FalckeThurley_PNAS2011}. The information content of frequency-encoding spiking increases strongly with decreasing CV of spike timing~\cite{Falcke2009_Chaos_Skupin}. A value of the coefficient of variation between 0.2 and 1.0 has been measured for the sequence of interspike intervals of intracellular Ca$^{2+}$ signaling~\cite{Falcke2008,falcke2014_SciSig,dragoni2011,sneyd2014,Lembong2016}. Hence,  CV is decreased compared with that of a Poisson process and even compared with that for first passage of a symmetric random walk. The experimental data are also compatible with the finding that the slower the recovery from negative feedback, the lower is CV~\cite{Falcke2008,falcke2014_SciSig,dragoni2011,sneyd2014,Lembong2016}. This strongly suggests that this ubiquitous cellular signaling system uses the mechanism of noise reduction described here to increase information transmission.

Optimal noise amplitudes minimize the CV of interspike intervals in noisy excitable systems~\cite{jung95,Pikovski97,haken93,engel2007,Schimansky2004d}, which has been termed coherence resonance. Our results define the conditions of optimal CV reduction in terms of system properties---an optimal number of steps from the inhibited state to the excitation threshold during slow recovery. At the same time, our results shed light on new aspects of coherence resonance, and  indeed may indicate that a more fundamental phenomenon underlies it. Since we believe excitable systems to be one of the most important applications of our results, we have chosen the term \emph{resonant} length.

Coming back to the widely used graphic example, what can the drunkard learn from our results? If he chooses a pub at the right distance from home, he will arrive home sober and relatively in time for breakfast.

\begin{acknowledgments}
VNF acknowledges support by IRTG 1740/TRP 2011/50151-0, funded by the DFG/FAPESP.
\end{acknowledgments}

\appendix

\section{The symmetric asymptotically Markovian case and its continuum limit}\label{sec:AppendixA}
The permanently symmetric case with a transient is defined by
\begin{eqnarray}
  f_{i,i\pm1}(t)&=&\lambda\left(1-e^{-\gamma t}\right).\label{eq:Mcase2}
\end{eqnarray}
Its continuum limit on a domain of length $L$ (with spatial coordinate $x$) satisfies
\begin{eqnarray}\label{eq:Contcase}
  \frac{\partial P(x,t)}{\partial t} &=& 2\lambda\left(1-e^{-\gamma t}\right) \frac{\partial^2 P(x,t)}{\partial x^2}.
\end{eqnarray}

The Fokker--Planck equation~\eqref{eq:Contcase} can be solved after applying a time transformation $t \rightarrow \tau$:
\begin{equation} \label{eq:Tau}
\tau(t)=2\int_0^t f_{i,i\pm1}(t')\,dt'=2\lambda\left(t+\frac{e^{-\gamma t}}{\gamma}-\frac{1}{\gamma}\right).
\end{equation}
The boundary and initial conditions, $P(L,t) = 0$, $\partial P(x,t)/\partial x|_{x=0}=0$, and $P(x,0)=\delta(x)$ specify the first-passage problem. We find
\begin{eqnarray} \label{eq:Pxtau}
P(x,\tau)=\frac{2}{L} \sum_{n=0}^{\infty} \cos \left(k_n x \right) e^{-k_n^2 \tau},
\end{eqnarray}
where $k_n = \pi (2n+1)/(2L)$. The $i$th moment of the first-passage time is given by
\begin{eqnarray} \label{eq:ithMom}
\langle t^i \rangle &=& \int_0^\infty t^i F(t) \,dt,\\
F(t) &=& - \frac{d}{dt} \int_0^L P(x,\tau(t)) \,dx.
\end{eqnarray}
With $a_n=2\lambda k_n^2/ \gamma$, we obtain
\begin{eqnarray} \label{eq:FirstMom}
\langle t \rangle = \frac{4}{\pi \gamma} \sum_{n=0}^{\infty} \frac{(-1)^n}{2n+1} e^{a_n} a_n^{-a_n} \Gamma \left(a_n,a_n\right),
\end{eqnarray}
where $\Gamma(y,x)$ is the lower incomplete gamma function. The second moment is
\begin{align} \label{eq:SecondMom}
\langle t^2 \rangle ={}&\frac{32 L^4}{\lambda^2 \pi^5} \sum_{n=0}^{\infty}  \frac{(-1)^n}{(2n+1)^5}  e^{a_n} \nonumber \\
&\times{}_2F_2 \left(\left\lbrace a_n,a_n \right\rbrace;\left\lbrace a_n+1,a_n+1 \right\rbrace;- a_n \right)
\end{align}
where ${}_2F_2$ is a hypergeometric function. Results for  CV are shown in Fig.~\ref{fig:CVsym}(a).  In the continuous case, CV does not exhibit a minimum in its dependence on the length $L$. Interestingly,  CV at small $L$ is very close to the minimum value of the discrete case.

The $a_n$ are small for $L \gg \lambda / \gamma$.  Therefore, in that limit, we can approximate the incomplete $\Gamma$ function and hypergeometric function by
\begin{gather} \label{eq:Approximations}
\Gamma \left(a_n,a_n\right) \approx a_n^{a_n-1} \ e^{-a_n},\nonumber\\
e^{a_n} \, {}_2F_2 \left(\left\lbrace a_n,a_n\right\rbrace;\left\lbrace a_n+1,a_n+1\right\rbrace;-a_n \right) \approx 1,\nonumber
\end{gather}
and find
\begin{eqnarray} \label{eq:CvLargeL}
\mathrm{CV}(L \gg \lambda/\gamma) \approx \sqrt{\frac{5}{3} \frac{L^4}{\lambda^2} \frac{\lambda^2}{L^4}-1} = \sqrt{\frac{2}{3}}\;,
\end{eqnarray}
i.e.,  CV approaches monotonically the value of a symmetric random walk with constant transition rates for large $L$.

\section{Numerical methods}\label{sec:AppendixB}

\emph{Evaluation of Eqs.~\eqref{eq:Iinh} and \eqref{eq:Marksolution}}: If $N$ is large and the ratio between the fastest state transition rate and the relaxation rate $\gamma$ is larger than $10^4$, high precision of the numerical calculations is required for the use of Eqs.~\eqref{eq:Iinh} and \eqref{eq:Marksolution}. \emph{A priori} known values like $\tilde{I}_{N-1,N}^0(s=0)=1$ can be used to monitor the precision of the calculations. The matrix products become very large at intermediate values of $j\gamma$ during the summation in Eqs.~\eqref{eq:Iinh} and \eqref{eq:Marksolution}, and their sign alternates such that two consecutive summands nearly cancel. Intermediate summands are of order larger than $10^{17}$, and thus we face loss of significant figures even with the numerical floating-point number format long double. We used Arb, a C numerical library for arbitrary-precision interval arithmetic~\cite{Johansson2017arb}, to circumvent this problem. It allows for arbitrary precision in calculations with  Eqs.~\eqref{eq:Iinh} and \eqref{eq:Marksolution}. Computational speed is the only limitation with this library and has determined the parameter range for which we established analytical results. We were able to go to a time-scale separation of $\approx10^6$ using this library.

\emph{Simulation method}: We use simulations for comparison with the analytic solutions and for systems with large $N$. The $\Psi_{l,j}(t,t-t')$ are functions only of $t$ for a given $t'$. Since $t'$ is the time when the process moved into state $l$, it is always known. The simulation algorithm generates in each iteration the cumulative distribution $\int_{t'}^t d\theta\, \sum_{i=1}^{N_\mathrm{out}^{(l)}}\Psi_{l,j_i}(\theta,\theta-t')$ and then draws the time $t_\mathrm{tr}$ for the next transition. The specific transition is chosen from the relative values $\Psi_{l,j_k}(t_\mathrm{tr},t_\mathrm{tr}-t')/\sum_{i=1}^{N_\mathrm{out}^{(l)}}\Psi_{l,j_i}(t_\mathrm{tr},t_\mathrm{tr}-t')$ by a second draw. We verified the simulation algorithm by comparison of simulated and calculated stationary probabilities [Eq.~\eqref{eq:nequation}], splitting probabilities at a variety of $t'$ values, simulations without recovery (``$\gamma=\infty$'') and the comparisons in Fig.~\ref{fig:TavScaling}. We used at least 20\,000 sample trajectories to calculate moments and up to 160\,000 to determine the location of the minima of  CV.

\section{The stationary probabilities}\label{sec:AppendixC}

The stationary probabilities are reached for $t\rightarrow\infty$. This entails $\Psi_{i,i\pm 1}(t,t-t')=\Psi_{i,i\pm 1}^\infty(t-t')=g_{i,i\pm 1}\left(t-t'\right)$. We start from the idea that the stationary probability $P_i$ of being in state $i$ is equal to the ratio of the total average time $T_i$ spent in $i$ divided by the total average time for large $t$:
\begin{eqnarray}\label{eq:StatProbGen}
P_i&=&\frac{T_i}{\sum_j^N T_j}.
\end{eqnarray}
$T_i$ is equal to the number $N_i$ of visits to state $i$ multiplied by the average dwell time $t_i$ in $i$:
\begin{eqnarray}
P_i&=&\frac{t_i N_i}{\sum_j^N t_j N_j}.
\end{eqnarray}
Each visit to state $i$ starts with a transition to $i$. The average number of transitions into $i$ is equal to the number of visits to its neighboring states multiplied by the probability that the transition out of the neighboring states is toward $i$. With the splitting probabilities $C_{i,i\pm1}(t=\infty)$ denoted by $C_{i,i\pm1}$, the $N_i$ obey
\begin{eqnarray}
  N_i &=& C_{i-1,i} N_{i-1} + C_{i+1,i} N_{i+1}.
\end{eqnarray}
Dividing by the total number of visits $\sum_{j=0}^N N_j$, we get
\begin{eqnarray}
  n_i &=& C_{i-1,i} n_{i-1} + C_{i+1,i} n_{i+1},\label{eq:nequation}\\
  1 &=&  \sum_{j=0}^N n_j,\label{eq:nequationnorm}\\
  P_i&=&\frac{t_i n_i}{\sum_{j=0}^N t_j n_j}.
\end{eqnarray}
We write Eq.~\eqref{eq:nequation} in matrix form, $\mathbb{M}\mathbb{n}=\mathbb{0}$, with
\begin{eqnarray}
  M_{i,i\pm 1} &=& -\tilde{g}_{i\pm 1,i}(0),\\
  M_{i,i} &=& 1,
\end{eqnarray}
and all other entries 0. We checked for $N=2,3,4$ that $\det\mathbb{M}=0$ holds. Equation~\eqref{eq:nequation} determines $\mathbb{n}$ only up to a common factor, which is then fixed by Eq.~\eqref{eq:nequationnorm}.

The average dwell times $t_i$ in the states are initially affected by the recovery from negative feedback, but are constant at large $t$. They have contributions $t_{i,i\pm1}$ from both transitions to $i\pm1$, with weights set by $C_{i,i\pm1}$:
\begin{eqnarray}
  t_i &=& C_{i,i+1} t_{i,i+1} + C_{i,i-1} t_{i,i-1}.
\end{eqnarray}
$C_{i,i\pm1} t_{i,i\pm1} = -(\partial/\partial s)\tilde{g}_{i,i\pm1}|_{s=0}$ holds owing to the normalization of the $\Psi_{i,i\pm1}$. This leads finally to
\begin{eqnarray}
  t_i &=& \left.-\frac{\partial}{\partial s} \left(\tilde{g}_{i,i-1} + \tilde{g}_{i,i+1}\right)\right|_{s=0}.\label{eq:Tequation}
\end{eqnarray}
To give a specific example, the stationary-state probabilities for $N=2$ are
\begin{align}
  \mathbb{P} = {}&\frac{1}{(\partial/\partial s)\left(C_{1,0}\tilde{g}_{0,1}+\tilde{g}_{1,0} + \tilde{g}_{1,2}+C_{1,2}\tilde{g}_{2,1}\right)}\nonumber\\[12pt]
  &\left.\times\begin{pmatrix}
    C_{1,0}(\partial/\partial s)\tilde{g}_{0,1} \\[4pt] (\partial/\partial s) \left(\tilde{g}_{1,0} + \tilde{g}_{1,2}\right)\\[4pt] C_{1,2}(\partial/\partial s)\tilde{g}_{2,1}
  \end{pmatrix}\right|_{s=0}.\nonumber\label{eq:StatProb}
\end{align}
\vspace{0.1cm}

%\bibliography{../../testbib}

%

\end{document}